\documentclass[fleqn,12pt]{article}

\usepackage{latexsym,ifthen,epsfig,color}
\usepackage{amsmath,amssymb,amsthm}
\usepackage{float}

\newcommand{\join}{\text{\textcircled{{\footnotesize 1}}}}
\newcommand{\cojoin}{\text{\textcircled{{\footnotesize 0}}}}

\newcommand{\NP}{\ensuremath{\mathbb{NP}}}

\newtheorem{clai}{Claim}[section]
\newtheorem{theo}{Theorem}
\newtheorem{lemma}{Lemma}

\newtheorem{coro}{Corollary}

\begin{document}

\author{
Andreas Brandst\"adt\\
\small Institut f\"ur Informatik, Universit\"at Rostock, D-18051 Rostock, Germany\\
\small \texttt{andreas.brandstaedt@uni-rostock.de}
}

\title{Finding Efficient Domination for $(S_{1,2,5},S_{3,3,3})$-Free Chordal Bipartite Graphs in Polynomial Time}

\maketitle

\begin{abstract}
A vertex set $D$ in a finite undirected graph $G$ is an {\em efficient dominating set} (\emph{e.d.s.}\ for short) of $G$ if every vertex of $G$ is dominated by exactly one vertex of $D$. The \emph{Efficient Domination} (ED) problem, which asks for the existence of an e.d.s.\ in $G$, is known to be \NP-complete for 
chordal bipartite graphs as well as for $P_7$-free graphs, and even for very restricted $H$-free bipartite graph classes such as for $K_{1,4}$-free bipartite graphs as well as for $C_4$-free bipartite graphs while it is solvable in polynomial time for $P_8$-free bipartite graphs as well as for $S_{1,3,3}$-free bipartite graphs and for $S_{1,1,5}$-free bipartite graphs. Here we show that ED can be solved in polynomial time for $(S_{1,2,5},S_{2,3,3})$-free chordal bipartite graphs. 
\end{abstract}


\section{Introduction}\label{sec:intro}

Let $G=(V,E)$ be a finite undirected graph. A vertex $v$ {\em dominates} itself and its neighbors. A vertex subset $D \subseteq V$ is an {\em efficient dominating set} ({\em e.d.s.}\ for short) of $G$ if every vertex of $G$ is dominated by exactly one vertex in $D$; for any e.d.s.\ $D$ of $G$, $|D \cap N[v]| = 1$ for every $v \in V$ (where $N[v]$ denotes the closed neighborhood of $v$).
Note that not every graph has an e.d.s.; the {\sc Efficient Dominating Set} ({\sc ED} for short) problem asks for the existence of an e.d.s.\ in a given graph~$G$.
The notion of efficient domination was introduced by Biggs \cite{Biggs1973} under the name {\em perfect code}.

\medskip

The Exact Cover Problem asks for a subset ${\cal F'}$ of a set family ${\cal F}$ over a ground set, say $V$, containing every vertex in $V$ exactly once, i.e., ${\cal F'}$ forms a partition of $V$. As shown by Karp \cite{Karp1972}, this problem is \NP-complete even for set families containing only $3$-element subsets of $V$ (see problem X3C [SP2] in \cite{GarJoh1979}).

\medskip

Clearly, ED is the Exact Cover problem for the closed neighborhood hypergraph of $G$, i.e., if $D=\{d_1,\ldots,d_k\}$ is an e.d.s.\ of $G$ then $N[d_1] \cup \ldots \cup N[d_k]$ forms a partition of $V$ (we call it the {\em e.d.s.\ property}). 
In \cite{BanBarSla1988,BanBarHosSla1996}, it was shown that the ED problem is \NP-complete. 

\medskip

For a set ${\cal F}$ of graphs, a graph $G$ is called {\em ${\cal F}$-free} if $G$ contains no induced subgraph isomorphic to a member of ${\cal F}$; in particular, we say that $G$ is {\em $H$-free} if $G$ is $\{H\}$-free. 

For $i \ge 1$, let $P_i$ denote the chordless path with $i$ vertices, and let $K_i$ denote the complete graph with $i$ vertices (clearly, $P_2=K_2$).
For $i \ge 4$, let $C_i$ denote the chordless cycle with $i$ vertices.

For indices $i,j,k \ge 0$, let $S_{i,j,k}$ denote the graph with vertices $u,r_1,\ldots,r_i$, $s_1,\ldots,s_j$, $t_1,\ldots,t_k$ (and {\em center} $u$) such that the subgraph induced by $u,r_1,\ldots,r_i$ forms a $P_{i+1}$ $(u,r_1,\ldots,r_i)$, the subgraph induced by $u,s_1,\ldots,s_j$ forms a $P_{j+1}$ $(u,s_1,\ldots,s_j)$, and the subgraph induced by $u,t_1,\ldots,t_k$ forms a $P_{k+1}$ $(u,t_1,\ldots,t_k)$, and there are no other edges in $S_{i,j,k}$. Thus, claw is $S_{1,1,1}$, chair is $S_{1,1,2}$, and $P_k$ is isomorphic to $S_{0,0,k-1}$. 

\medskip

In \cite{EscWan2014,SmaSla1995,YenLee1996}, it was shown that ED is \NP-complete for $2P_3$-free chordal unipolar graphs and thus, in general, for $P_7$-free graphs.
In \cite{BraMos2016}, ED is solvable in polynomial time for $P_6$-free graphs (which leads to a dichotomy). 

\medskip

A bipartite graph $G$ is {\em chordal bipartite} if $G$ is $C_{2k}$-free for every $k \ge 3$. 
Lu and Tang \cite{LuTan2002} showed that ED is \NP-complete for chordal bipartite graphs (i.e., hole-free bipartite graphs). Thus, for every $k \ge 3$, ED is \NP-complete for $C_{2k}$-free bipartite graphs.
Moreover, ED is \NP-complete for planar bipartite graphs \cite{LuTan2002} and even for planar bipartite graphs of maximum degree 3 \cite{BraMilNev2013} and girth at least $g$ for every fixed $g$ \cite{Nevri2014}. Thus, ED is \NP-complete for $K_{1,4}$-free bipartite graphs and for $C_4$-free bipartite graphs.

\medskip

In \cite{BraFicLeiMil2015}, it is shown that one can extend polynomial time algorithms for Efficient Domination to such algorithms for weighted Efficient Domination. Thus, from now on, we focus on the unweighted ED problem.

\medskip

In \cite{BraLeiRau2012}, it is shown that ED is solvable in polynomial time for AT-free graphs. Moreover,
in \cite{BraLeiRau2012}, it is shown that ED is solvable in polynomial time for interval bigraphs, and convex bipartite graphs are a subclass of them (and of chordal 
bipartite graphs). Moreover, Lu and Tang \cite{LuTan2002} showed that ED is solvable in linear time for bipartite permutation graphs (which is a subclass of convex bipartite graphs). 

\medskip

In \cite{BraMos2019} we have shown that ED is solvable in polynomial time for $P_7$-free bipartite graphs as well as for $S_{2,2,4}$-free bipartite graphs. 
Moreover, we have shown that ED is solvable in polynomial time for $P_8$-free bipartite graphs \cite{BraMos2021/3} as well as for $S_{1,1,5}$-free bipartite graphs \cite{BraMos2021/2} and for $S_{1,3,3}$-free bipartite graphs \cite{BraMos2021/1}. 
Now in this manuscript, we show: 

\begin{theo}\label{EDS125S333frchordbippoltime}
For $(S_{1,2,5},S_{3,3,3})$-free chordal bipartite graphs, the ED problem is solvable in polynomial time. 
\end{theo} 

\section{Preliminaries}

Recall that $G=(X,Y,E)$ is an $(S_{1,2,5},S_{3,3,3})$-free chordal bipartite graph, say, every vertex in $X$ is black, and every vertex in $Y$ is white; 
let $V = X \cup Y$. 
A vertex $u \in V$ is {\em forced} if $u \in D$ for every e.d.s.\ $D$ of $G$; $u$ is {\em excluded} if $u \notin D$ for every e.d.s.\ $D$ of $G$.
By a forced vertex, $G$ can be reduced to $G'$ as follows:

\begin{clai}\label{forcedreduction}
If $u$ is forced then $G$ has an e.d.s.\ $D$ with $u \in D$ if and only if the reduced graph $G'=G \setminus N[u]$ has an e.d.s.\ $D'=D \setminus \{u\}$ such that all vertices in $N^2(u)$ are excluded in $G'$.
\end{clai}

Analogously, if we assume that $v \in D$ for a vertex $v \in V=X \cup Y$ then $u \in V$ is {\em $v$-forced} if $u \in D$ for every e.d.s.\ $D$ of $G$ with $v \in D$,
and $u$ is {\em $v$-excluded} if $u \notin D$ for every e.d.s.\ $D$ of $G$ with $v \in D$. For checking whether $G$ has an e.d.s.\ $D$ with $v \in D$, we can clearly reduce $G$ by forced vertices as well as by $v$-forced vertices when we assume that $v \in D$:

\begin{clai}\label{vforcedreduction}
If we assume that $v \in D$ and $u$ is $v$-forced then $G$ has an e.d.s.\ $D$ with $v \in D$ if and only if the reduced graph $G'=G \setminus N[u]$ has an e.d.s.\ $D'=D \setminus \{u\}$ with $v \in D'$ such that all vertices in $N^2(u)$ are $v$-excluded in $G'$.
\end{clai}

Similarly, for $k \ge 2$, $u \in V$ is {\em $(v_1,\ldots,v_k)$-forced} if $u \in D$ for every e.d.s.\ $D$ of $G$ with $v_1,\ldots,v_k \in D$, and correspondingly, $u \in V$ is {\em $(v_1,\ldots,v_k)$-excluded} if $u \notin D$ for such e.d.s.\ $D$, and $G$ can be reduced by the same principle.

\medskip

Clearly, for every component of $G$, the e.d.s.\ problem is independently solvable. Thus, we can assume that $G$ is connected.   

\medskip

Let $dist_G(u,v)$ denote the minimum distance, i.e., the minimum length of a path between $u$ and $v$ in $G$. By the e.d.s.\ property, the distance between two 
$D$-vertices is at least $3$. Moreover, if $u,u' \in D \cap X$ or $u,u' \in D \cap Y$ then $dist_G(u,u') \ge 4$.  

\medskip

If for an e.d.s.\ $D$ in the bipartite graph $G=(X,Y,E)$, $|D|=1$, say without loss of generality, $D =\{x\}$ with $x \in X$ and 
$D \cap Y=\emptyset$, then every $y \in Y$ must have the $D$-neighbor $x \in D$, i.e., $x \join Y$ and by the e.d.s.\ property, $|X|=1$, i.e., $X=\{x\}$ (else there is no such e.d.s.\ in $G$), which is a trivial e.d.s.\ solution. 

\medskip

Now assume that $|D| \ge 2$, and without loss of generality, $D \cap X \neq \emptyset$. 
Since $G$ is connected, every $x \in D \cap X$ must have at least distance $2$ in $G$, say $(x,y,x')$ induce a $P_3$ in $G$. Then by the e.d.s.\ property, 
$y,x' \notin D$ and $x'$ must have a $D$-neighbor $y' \in D \cap Y$ (else there is no such e.d.s.\ in $G$). Thus, $D \cap X \neq \emptyset$ and 
$D \cap Y \neq \emptyset$.  

\medskip

If $G$ is $P_7$-free bipartite then by \cite{BraMos2019}, the e.d.s.\ problem is solvable in polynomial time. 
Moreover, if $G$ is $P_8$-free bipartite then by \cite{BraMos2021/3}, the e.d.s.\ problem is solvable in polynomial time. 
Now assume that there are $P_8$'s in $G$ (but no such $S_{1,2,5}$'s and $S_{3,3,3}$'s in $G$); let $P=(u_1,v_1,\ldots,u_4,v_4)$ induce a $P_8$ in $G$ 
(possibly $u_i \in X$, $v_j \in Y$ or $u_i \in Y$, $v_j \in X$). 
Clearly, by the e.d.s.\ property, there are at most three such $D$-vertices in $P_8$ $P$. 

\begin{lemma}\label{P5u1v1u2v2u3u2inD}
If $P=(u_1,v_1,u_2,v_2,u_3)$ induce a $P_5$ in $G$ with midpoint $u_2 \in D$ then $u_2$ is midpoint of $P_7$ $P'=(v,u_1,v_1,u_2,v_2,u_3,v')$, and $v \in D$ as well as $v' \in D$ are $u_2$-forced. 
\end{lemma}

\noindent
{\em Proof.}
Let $P=(u_1,v_1,u_2,v_2,u_3)$ induce a $P_5$ in $G$ with midpoint $u_2 \in D$. Then by the e.d.s.\ property, $v_1,v_2,u_1,u_3 \notin D$ and 
$u_1,u_3$ must have $D$-neighbors, say $u_1v \in E$ with $v \in D$. 

If $N(u_1) \join \{u_2\}$ then there is no such e.d.s.\ in $G$ with $u_2 \in D$. Analogously, if $N(u_3) \join \{u_2\}$ then there is no such e.d.s.\ in $G$ with $u_2 \in D$. Now assume that $u_1v \in E$ with $vu_2 \notin E$.  

Since $G$ is $C_6$-free, i.e., $(v,u_1,v_1,u_2,v_2,u_3)$ does not induce a $C_6$ in $G$, we have $u_3v \notin E$, i.e., $u_3v' \in E$ with $v' \in D$. Then $u_2$ is midpoint of $P_7$ $P'=(v,u_1,v_1,u_2,v_2,u_3,v')$. 

\medskip

If $N(u_1)=\{v,v_1\}$ then $v \in D$ is $u_2$-forced. Now assume that $|N(u_1)| \ge 3$, say $v_0u_1 \in E$ and $vu_1 \in E$ with $v \in D$. 
If $(N(u_1) \setminus \{v\}) \join \{u_2\}$ then again $v \in D$ is $u_2$-forced. Now assume that $v_0u_2 \notin E$. 
Then, since $(v_0,u_1,v_1,u_2,v_2,u_3)$ does not induce a $C_6$ in $G$, we have $v_0u_3 \notin E$, and $v_0$ must have a $D$-neighbor $u \in D$, i.e., $v_0u \in E$. 
Clearly, by the e.d.s.\ property, $uv_1 \notin E$, $uv_2 \notin E$, $uv \notin E$ and $uv' \notin E$. 

But then $(u_1,v,v_0,u,v_1,u_2,v_2,u_3,v')$ (with midpoint $u_1$) induce an $S_{1,2,5}$ in $G$, which is a contradiction. 
Thus, $(N(u_1) \setminus \{v\}) \join \{u_2\}$, i.e., every other neighbor of $u_1$ contacts $u_2$, and $v \in D$ is $u_2$-forced. 

\medskip

Analogously, since $v' \in D$ with $u_3v' \in E$, $(N(u_3) \setminus \{v'\}) \join \{u_2\}$, i.e., every other neighbor of $u_3$ contacts $u_2$, and $v' \in D$ is $u_2$-forced. Thus, Lemma \ref{P5u1v1u2v2u3u2inD} is shown. 
\qed

\begin{lemma}\label{P5u1u2u3vinD}
If $P=(u_1,v_1,u_2,v_2,u_3)$ induce a $P_5$ in $G$, $D \cap V(P)=\emptyset$ and $u_1,u_3$ have a common $D$-neighbor $v \in D$, then $u_2v \in E$ and
$v_1,v_2$ do not have any common $D$-neighbor. 
\end{lemma}

\noindent
{\em Proof.}
Let $P=(u_1,v_1,u_2,v_2,u_3)$ induce a $P_5$ in $G$, $D \cap V(P)=\emptyset$ and $u_1,u_3$ have a common $D$-neighbor $v \in D$. Since $G$ is $C_6$-free, i.e., 
$(v,u_1,v_1,u_2,v_2,u_3)$ does not induce a $C_6$ in $G$, we have $u_2v \in E$. 

Moreover, by the e.d.s.\ property, $v_i \notin D$, $i \in \{1,2\}$, and $v_i$ must have a $D$-neighbor, say $uv_1 \in E$ with $u \in D$. 
Since $(u,v_1,u_1,v,u_3,v_2)$ does not induce a $C_6$ in $G$, we have $uv_2 \notin E$, i.e., 
$u'v_2 \in E$ with $u' \in D$, $u \neq u'$, and $u'v_1 \notin E$, i.e., $v_1,v_2$ do not have any common $D$-neighbor. 
\qed 

\begin{coro}\label{P7u1u2u3u4vinD}
If $P=(u_1,v_1,u_2,v_2,u_3,v_3,u_4)$ induce a $P_7$ in $G$, $D \cap V(P)=\emptyset$ and $u_1,u_2,u_3,u_4$ have a common $D$-neighbor $v \in D$, i.e., $vu_i \in E$, $1 \le i \le 4$, then $v_1,v_2,v_3$ do not have any common $D$-neighbor. 
\end{coro}

\noindent
{\em Proof.}
Let $P=(u_1,v_1,u_2,v_2,u_3,v_3,u_4)$ induce a $P_7$ in $G$ and $u_1,u_2,u_3,u_4$ have a common $D$-neighbor $v \in D$. 
By the e.d.s.\ property, $v_1,v_2,v_3 \notin D$ and $v_1,v_2,v_3$ must have $D$-neighbors in $G$. Let $uv_1 \in E$ with $u \in D$. 
Since $G$ is $C_6$-free, i.e., $(u,v_1,u_1,v,u_4,v_3)$ does not induce a $C_6$ in $G$, $uv_3 \notin E$, i.e., $v_1,v_3$ do not have any common $D$-neighbor.

Recall that by Lemma \ref{P5u1u2u3vinD}, for $P_5$ $(u_1,v_1,u_2,v_2,u_3)$, $v_1$ and $v_2$ do not have any common $D$-neighbor. Moreover,   
for $P_5$ $(u_2,v_2,u_3,v_3,u_4)$, $v_2$ and $v_3$ do not have any common $D$-neighbor, i.e., $v_1,v_2,v_3$ do not have any common $D$-neighbor.
\qed

\section{Distance levels $N_i$, $i \ge 0$}

Recall that $G=(X,Y,E)$ is an $(S_{1,2,5},S_{3,3,3})$-free chordal bipartite graph and let $V=X \cup Y$.  
Let $N_0:=D_{basis}$, and for every $D_{basis}$-forced vertex $d \in D$, $d \in D_{basis}$, i.e., $D_{basis}:= D_{basis} \cup \{d\}$. 
Let $N_i$, $i \ge 1$, be the distance levels of $D_{basis}$. 

Recall that $D \cap X \neq \emptyset$ and $D \cap Y \neq \emptyset$; assume that $N_0 \cap X \neq \emptyset$ and $N_0 \cap Y \neq \emptyset$, say $x \in N_0 \cap X$ and $y \in N_0 \cap Y$ with $dist_G(x,y) = 3$, say $(x,y_1,x_1,y)$ induce a $P_4$ in $G$ with $y_1,x_1 \in N_1$, and every $(x,y)$-forced vertex is in $D_{basis}$. 
By the e.d.s.\ property, we have: 
\begin{equation}\label{DN1N2empty}
D \cap (N_1 \cup N_2)=\emptyset. 
\end{equation}

If there is a $u \in N_2$ with $N(u) \cap N_3=\emptyset$ then by the e.d.s.\ property, there is no such e.d.s.\ $D$ in $G$ with $N_0=D_{basis}$. Thus, we assume:
\begin{equation}\label{uinN2neighbinN3}
\mbox{ Every vertex } u \in N_2 \mbox{ has a $D$-neighbor in } N_3.
\end{equation}
If $|N(u) \cap N_3| = 1$, say $N(u) \cap N_3=\{v\}$ then $v$ is $D_{basis}$-forced, and one can update $D_{basis}:= D_{basis} \cup \{v\}$ and redefine the distance levels $N_i$, $i \ge 1$, with respect to $D_{basis}$ as above. Now assume:
\begin{equation}\label{uinN2twoneighbinN3}
\mbox{ For every } u \in N_2, |N(u) \cap N_3| \ge 2.
\end{equation}
If for $v \in N_3$, $N(v) \cap (N_3 \cup N_4)=\emptyset$ then by (\ref{DN1N2empty}) and the e.d.s.\ property, $v$ is $D_{basis}$-forced. 
Moreover, if there are two such $v_i \in N_3$, $N(v_i) \cap (N_3 \cup N_4)=\emptyset$, $i \in \{1,2\}$, with common neighbor $u \in N_2$, i.e., 
$uv_i \in E$, $i \in \{1,2\}$, then by the e.d.s.\ property, there is no such e.d.s.\ $D$ in $G$ with $N_0=D_{basis}$. 
Thus assume that such $D_{basis}$-forced vertices $v_i \in N_3$, $i \in \{1,2\}$, do not have any common neighbor in $N_2$. 

Then for every $v \in N_3$ with $N(v) \cap (N_3 \cup N_4)=\emptyset$, one can update $D_{basis}:= D_{basis} \cup \{v\}$ and redefine the distance levels 
$N_i$, $i \ge 1$, with respect to $D_{basis}$ as above. 
Thus, we assume:
\begin{equation}\label{vinN3neighbinN3N4}
\mbox{ Every } v \in N_3 \mbox{ has a neighbor in } N_3 \cup N_4.
\end{equation}

In particular, we have:

\begin{lemma}\label{uinN2endpointnoP6N0N1}
For every $r_2 \in N_2$, $r_2$ is no endpoint of a $P_6$ whose remaining vertices are in $N_0 \cup N_1$.
\end{lemma}

\noindent
{\bf Proof.} 
Let $r_2 \in N_2$ be the endpoint of a $P_k$ $(r_2,r_1,v_1,\ldots,v_k)$, $1 \le k$, with $r_1 \in N_1$ and $v_1,\ldots,v_k \in N_0 \cup N_1$. 
Clearly, $(r_2,r_1,r_0)$ induce a $P_3$ in $G$ with $r_1 \in N_1$ and $r_0 \in N_0$. 

\medskip

Suppose to the contrary that $r_2$ is the endpoint of a $P_6$ $(r_2,r_1,v_1,v_2,v_3,v_4)$ with $r_1 \in N_1$ and $v_1,\ldots,v_4 \in N_0 \cup N_1$.
By (\ref{uinN2twoneighbinN3}), $r_2$ must have at least two neighbors in $N_3$, say $r_2r_3 \in E$, $r_2r'_3 \in E$ with $r_3,r'_3 \in N_3$ and without loss of generality, $r_3 \in D$. By the e.d.s.\ property, $r'_3 \notin D$ and $r'_3$ must have a $D$-neighbor $s \in D \cap (N_3 \cup N_4)$. 
But then $(r_2,r_3,r'_3,s,r_1,v_1,v_2,v_3,v_4)$ (with midpoint $r_2$) induce an $S_{1,2,5}$, which is a contradiction.

Thus, $r_2$ is no endpoint of a $P_6$ whose remaining vertices are in $N_0 \cup N_1$, i.e., $r_2$ is the endpoint of a $P_5$ $(r_2,r_1,v_1,v_2,v_3)$ with $r_1 \in N_1$
and $v_1,v_2,v_3 \in N_0 \cup N_1$ or a $P_4$ $(r_2,r_1,v_1,v_2)$ with $r_1 \in N_1$ and $v_1,v_2 \in N_0 \cup N_1$ 
or a $P_3$ $(r_2,r_1,r_0)$ with $r_1 \in N_1$ and $r_0 \in N_0$, and Lemma \ref{uinN2endpointnoP6N0N1} is shown.
\qed

\begin{lemma}\label{N6emptyN5indep}
Every $r_2 \in N_2$ is no endpoint of a $P_5$ $P=(r_2,r_3,r_4,r_5,r_6)$ with $r_3 \in N_3$, $r_4 \in N_3 \cup N_4$, $r_5 \in N_3 \cup N_4 \cup N_5$, 
$r_6 \in N_3 \cup N_4 \cup N_5 \cup N_6$, i.e., $N_6=\emptyset$, $N_5$ is independent, every edge in $N_4$ does not contact $N_5$, and for every edge $r_3s_3$ in $N_3$, there is no such $P_4$ $(r_3,s_3,r_4,r_5)$ with $r_4 \in N_4$ and $r_5 \in N_4 \cup N_5$.    
\end{lemma}

\noindent
{\bf Proof.}
Let $(x,y_1,x_1,y)$ induce a $P_4$ in $N_0 \cup N_1$ with $x,y \in N_0$, $y_1,x_1 \in N_1$, and every $(x,y)$-forced vertex is in $N_0=D_{basis}$. 

\medskip

Suppose to the contrary that there is an $r_2 \in N_2$ with endpoint of a $P_5$ $P=(r_2,r_3,r_4,r_5,r_6)$, $r_3 \in N_3$, $r_4 \in N_3 \cup N_4$, $r_5 \in N_3 \cup N_4 \cup N_5$, $r_6 \in N_3 \cup N_4 \cup N_5 \cup N_6$. 

\medskip

First suppose to the contrary that $N_6 \neq \emptyset$, say $(r_2,r_3,r_4,r_5,r_6)$ induce a $P_5$ in $G$ with $r_i \in N_i$, $2 \le i \le 6$. 
Without loss of generality, assume that $r_2 \in X$. 

If $r_2y_1 \in E$ then $(y_1,x,x_1,y,r_2,r_3,r_4,r_5,r_6)$ (with midpoint $y_1$) induce an $S_{1,2,5}$ in $G$, which is a contradiction. Thus, $r_2y_1 \notin E$ and $r_1r_2 \in E$ with $r_1 \in N_1$, $r_1 \neq y_1$. 

\medskip

First assume that $xr_1 \in E$. If $r_1x_1 \in E$ then $(r_1,x,x_1,y,r_2,r_3,r_4,r_5,r_6)$ (with midpoint $r_1$) induce an $S_{1,2,5}$ in $G$, which is a contradiction. 
Thus, $r_1x_1 \notin E$. But then $r_2$ is the endpoint of a $P_6$ $(r_2,r_1,x,y_1,x_1,y)$ with $x_1,r_1,y_1 \in N_1$ and $x,y \in N_0$, which is a contradiction 
by Lemma \ref{uinN2endpointnoP6N0N1}. Thus, $xr_1 \notin E$.

\medskip

Next assume that $r_0r_1 \in E$ with $r_0 \in N_0$, $x \neq r_0$. 
Since $G[N_0 \cup N_1]$ is connected, $r_0$ and $y$ must have distance at least $3$; without loss of generality, $dist_G(r_0,y)=3$, say $(r_0,r'_1,x'_1,y)$ induce a $P_4$ in $G[N_0 \cup N_1]$ (possibly $r_1=r'_1$ or $x_1=x'_1$).

If $r_1=r'_1$ and $x_1=x'_1$, i.e., $r_1x_1 \in E$, then again $(r_1,r_0,x_1,y,r_2,r_3,r_4,r_5,r_6)$ (with midpoint $r_1$) induce an $S_{1,2,5}$ in $G$, which is a contradiction. Thus, either $r_1 \neq r'_1$ or $x_1 \neq x'_1$, say without loss of generality, $r_1 \neq r'_1$. Since by Lemma~\ref{uinN2endpointnoP6N0N1}, $r_2$ is no such endpoint of a $P_6$ $(r_2,r_1,r_0,r'_1,x'_1,y)$, $r'_1r_2 \in E$. But then again $(r'_1,r_0,x'_1,y,r_2,r_3,r_4,r_5,r_6)$ (with midpoint $r'_1$) induce an $S_{1,2,5}$ in $G$, which is a contradiction. Thus, $N_6=\emptyset$. 

\medskip

Next suppose to the contrary that $N_5$ is not independent, say $(r_2,r_3,r_4,r_5,s_5)$ induce a $P_5$ in $G$ with $r_i \in N_i$, $2 \le i \le 5$ and $s_5 \in N_5$. But then again there are such $S_{1,2,5}$'s in $G$, which is a contradiction. Thus, $N_5$ is independent.

\medskip 

Next suppose to the contrary that an edge $r_4s_4 \in N_4$ contacts $N_5$. Without loss of generality, $s_4r_5 \in E$ with $r_5 \in N_5$,  
i.e., $(r_2,r_3,r_4)$ induce a $P_3$ in $G$ with $r_i \in N_i$, $2 \le i \le 4$, say $(r_2,r_3,r_4,s_4,r_5)$ induce a $P_5$ in $G$ with $r_i \in N_i$, 
$2 \le i \le 5$, and $s_4 \in N_4$. 
But then again there are such $S_{1,2,5}$'s in $G$, which is a contradiction. Thus, every edge in $N_4$ does not contact $N_5$. 

\medskip 

Finally suppose to the contrary that there is a $P_4$ $(r_3,s_3,r_4,r_5)$ with edge $r_3s_3$ in $N_3$, $r_4 \in N_4$ and $r_5 \in N_4 \cup N_5$. 
But then again there are such $S_{1,2,5}$'s in $G$, which is a contradiction. Thus, there is no such $P_4$ $(r_3,s_3,r_4,r_5)$ with $r_4 \in N_4$,  
$r_5 \in N_4 \cup N_5$, and Lemma~\ref{N6emptyN5indep} is shown.  
\qed 

\begin{coro}\label{coroN6emptyN5indep}
For every $r_2 \in N_2$ and $r_3 \in D \cap N_3$ with $r_2r_3 \in E$, $r_2 \in N_2$ is no endpoint of $P_4$ $(r_2,r_3,r_4,r_5)$ with $r_3 \in D \cap N_3$, 
$r_4 \in N_3 \cup N_4$, $r_5 \in N_3 \cup N_4 \cup N_5$. 
\end{coro}

\begin{lemma}\label{2P3N3commonN2neighbor} 
For every $2P_3$ $(r_2,r_3,r_4)$, $(s_2,s_3,s_4)$ in $G$ with $r_3,s_3 \in D \cap N_3 \cap X$ or $r_3,s_3 \in D \cap N_3 \cap Y$, $r_2,s_2 \in N_2$, 
$r_4,s_4 \in N_3 \cup N_4$, $r_2$ and $s_2$ must have a common neighbor in $N_1$.  
\end{lemma}

\noindent
{\em Proof.}
Let $(r_2,r_3,r_4)$, $(s_2,s_3,s_4)$ be $P_3$'s with $r_3,s_3 \in D \cap N_3 \cap X$ or $r_3,s_3 \in D \cap N_3 \cap Y$, $r_2,s_2 \in N_2$, $r_4,s_4 \in N_3 \cup N_4$. By the e.d.s.\ property, $(r_2,r_3,r_4)$, $(s_2,s_3,s_4)$ induce a $2P_3$ in $G$. 

\medskip

Suppose to the contrary that $r_2$ and $s_2$ do not have any common neighbor in $N_1$, say $r_1r_2 \in E$, $s_1s_2 \in E$, $r_1,s_1 \in N_1$, $r_1 \neq s_1$. 

\medskip

First assume that $r_1$ and $s_1$ have a common neighbor $r_0 \in N_0$, i.e., $r_0r_1 \in $ and $r_0s_1 \in E$. Recall that by $(\ref{uinN2neighbinN3})$,
$r_2$ must have a $D$-neighbor in $N_3$, by (\ref{uinN2twoneighbinN3}), $|N(r_2)| \ge 2$, say $r_2r_3 \in E$ with $r_3 \in D \cap N_3$, 
$r_2r'_3 \in E$, and by the e.d.s.\ property, $r'_3$ must have a $D$-neighbor $r'_4 \in D \cap (N_3 \cup N_4)$. 
Since there is no such $C_6$ $(r_0,r_1,r_2,r'_3,s_2,s_1)$ in $G$, $r'_3s_2 \notin E$.   
But then $(r_2,r_3,r'_3,r'_4,r_1,r_0,s_1,s_2,s_3)$ (with midpoint $r_2$) induce an $S_{1,2,5}$ in $G$, which is a contradiction. 
Thus, $r_1$ and $s_1$ do not have any common neighbor in $N_0$, i.e., for $r_0,s_0 \in N_0$, $r_0 \neq s_0$, $(r_0,r_1,r_2,r_3,r_4)$, $(s_0,s_1,s_2,s_3,s_4)$,  
induce a $2P_5$ in $G$. 

\medskip

Then without loss of generality, $dist_G(r_0,s_0)=4$, say $(r_0,r'_1,t_1,s'_1,s_0)$ induce a $P_5$ in $N_0 \cup N_1$ (possibly $r'_1=r_1$ or $s'_1=s_1$). 
Then $t_0t_1 \in E$ with $t_0 \in N_0$.

If $r_1t_1 \in E$ and $s_1t_1 \in E$ then $(r_1,r_0,r_2,r_3,t_1,s_1,s_2,s_3,s_4)$ (with midpoint $r_1$) induce an $S_{1,2,5}$ in $G$, which is a contradiction. 
Thus, $r_1,s_1$ do not have any common $N_1$-neighbor, say without loss of generality, $r_1t_1 \notin E$. 
Since by Lemma~\ref{uinN2endpointnoP6N0N1}, $r_2$ is no endpoint of a $P_6$ $(r_2,r_1,r_0,r'_1,t_1,t_0)$, $r'_1r_2 \in E$.

\medskip

If $t_1s_1 \in E$ then $(s_1,s_0,s_2,s_3,t_1,r'_1,r_2,r_3,r_4)$ (with midpoint $s_1$) induce an $S_{1,2,5}$ in $G$, which is a contradiction.   
Thus, $t_1s_1 \notin E$, i.e., $t_1s'_1 \in E$. 
Since by Lemma~\ref{uinN2endpointnoP6N0N1}, $s_2$ is no endpoint of a $P_6$ $(s_2,s_1,s_0,s'_1,t_1,t_0)$, $s'_1s_2 \in E$.

But then $(s'_1,s_0,s_2,s_3,t_1,r'_1,r_2,r_3,r_4)$ (with midpoint $s'_1$) induce an $S_{1,2,5}$ in $G$, which is a contradiction.  

\medskip

Thus, $r_2$ and $s_2$ must have a common neighbor in $N_1$, and Lemma \ref{2P3N3commonN2neighbor} is shown.
\qed   

\begin{lemma}\label{atmosttwoDN3XorY} 
There are at most two $D$-vertices $r_3,s_3 \in D \cap N_3 \cap X$, and there are at most two $D$-vertices $r_3,s_3 \in D \cap N_3 \cap Y$. 
\end{lemma}

\noindent
{\em Proof.}
Recall that by (\ref{vinN3neighbinN3N4}), for $r_3 \in D \cap N_3$, $(r_2,r_3,r_4)$ induce a $P_3$ in $G$ with $r_2 \in N_2$, $r_4 \in N_3 \cup N_4$. 
Without loss of generality, assume that there are two $D$-vertices $r_3,s_3 \in D \cap N_3 \cap X$, say $(s_2,s_3,s_4)$ induce a $P_3$ in $G$ with $s_2 \in N_2$, $s_4 \in N_3 \cup N_4$, i.e., $(r_2,r_3,r_4)$, $(s_2,s_3,s_4)$ induce a $2P_3$ in $G$.  

\medskip

Suppose to the contrary that there are three $D$-vertices $r_3,s_3,t_3 \in D \cap N_3 \cap X$, say $(r_2,r_3,r_4)$, 
$(s_2,s_3,s_4)$, and $(t_2,t_3,t_4)$ induce $P_3$'s in $G$ with $r_2,s_2,t_2 \in N_2$ and $r_4,s_4,t_4 \in N_3 \cup N_4$. 
Then by the e.d.s.\ property, $(r_2,r_3,r_4)$, $(s_2,s_3,s_4)$, $(t_2,t_3,t_4)$ induce a $3P_3$ in $G$. 
Recall that by Lemma \ref{2P3N3commonN2neighbor}, there are common $N_1$-neighbors for $r_2$ and $s_2$, for $s_2$ and $t_2$ and for  
$r_2$ and $t_2$.
        
\medskip

If $r_2,s_2,t_2$ have a common neighbor in $N_1$, say without loss of generality, $r_1r_2 \in E$, $r_1s_2 \in E$ and $r_1t_2 \in E$ with $r_1 \in N_1$ then $(r_1,r_2,r_3,r_4,s_2,s_3,s_4,t_2,t_3,t_4)$ (with midpoint $r_1$) induce an $S_{3,3,3}$ in $G$, which is a contradiction. 
Thus, $r_2,s_2,t_2$ do not have any common neighbor in $N_1$. 

\medskip

Now assume that for $r_1,s_1,t_1 \in N_1$, $r_1r_2 \in E$, $r_1s_2 \in E$, $r_1t_2 \notin E$, $s_1s_2 \in E$, $s_1t_2 \in E$, $s_1r_2 \notin E$, 
and $t_1t_2 \in E$, $t_1r_2 \in E$, $t_1s_2 \notin E$.  
But then $(r_2,r_1,s_2,s_1,t_2,t_1)$ induce a $C_6$ in $G$, which is a contradiction.

\medskip

Thus, there are at most two $D$-vertices $r_3,s_3 \in D \cap N_3 \cap X$. Analogously, there are at most two $D$-vertices $r_3,s_3 \in D \cap N_3 \cap Y$, and 
Lemma \ref{atmosttwoDN3XorY} is shown.
\qed

\begin{lemma}\label{N5noDmidpointP5}
There are no $D$-midpoints in $D \cap N_5$ of a $P_5$ in $G$.
\end{lemma}

\noindent
{\bf Proof.} 
Let $(x,y_1,x_1,y)$ induce a $P_4$ in $G$ with $x,y \in N_0=D_{basis}$, $y_1,x_1 \in N_1$, and every $(x,y)$-forced vertex is in $N_0=D_{basis}$. 
Assume that $r_5 \in D \cap N_5$. Recall that by Lemma~\ref{N6emptyN5indep}, $N_6=\emptyset$, $N_5$ is independent, and every edge in $N_4$ does not contact $N_5$. 

\medskip

Suppose to the contrary that $r_5 \in D$ is a $D$-midpoint of a $P_5$ in $G$. Then $r_5$ does not contact vertices in $N_5$ and edges in $N_4$. 

\medskip

First assume that $(s_5,r_4,r_5,r'_4,s'_5)$ induce a $P_5$ in $G$ with midpoint $r_5 \in D$, $r_4,r'_4 \in N_4$, $s_5,s'_5 \in N_5$. 
Then by the e.d.s.\ property, $s_5,s'_5 \notin D$ and $s_5,s'_5$ must have $D$-neighbors $s_4,s'_4 \in D \cap N_4$, $s_4 \neq s'_4$ (else there is a $C_6$ $(s_4,s_5,r_4,r_5,r'_4,s'_5)$ in $G$ if $s_4=s'_4$). Let $s_3s_4 \in E$ with $s_3 \in N_3$ and $s_2r_3 \in E$ with $s_2 \in N_2$. 
Since by Lemma~\ref{N6emptyN5indep}, $s_2$ is no such $P_5$-endpoint $(s_2,s_3,s_4,s_5,r_4)$, $s_3r_4 \in E$. 
Moreover, since by Lemma~\ref{N6emptyN5indep}, $s_2$ is no such $P_5$-endpoint $(s_2,s_3,r_4,r_5,r'_4)$, $s_3r'_4 \in E$.
By the e.d.s.\ property, $s_3s'_4 \notin E$ but then $s_2$ is a $P_5$-endpoint of a $P_5$ $(s_2,s_3,r'_4,s'_5,s'_4)$, which is a contradiction 
by Lemma~\ref{N6emptyN5indep}. Thus, there is no such $P_5$ $(s_5,r_4,r_5,r'_4,s'_5)$ in $G$ with midpoint $r_5 \in D$, $r_4,r'_4 \in N_4$, $s_5,s'_5 \in N_5$. 

\medskip

Next assume that $(s_5,r_4,r_5,r'_4,r'_3)$ induce a $P_5$ in $G$ with midpoint $r_5 \in D$, $r_4,r'_4 \in N_4$, $s_5 \in N_5$, $r'_3 \in N_3$. 
Clearly, $r'_3r_4 \notin E$. Let $r'_2r'_3 \in E$ with $r'_2 \in N_2$. 
But then $r'_2$ is a $P_5$-endpoint of a $P_5$ $(r'_2,r'_3,r'_4,r_5,r_4)$, which is a contradiction by Lemma~\ref{N6emptyN5indep}.
Thus, there is no such $P_5$ $(s_5,r_4,r_5,r'_4,r'_3)$ in $G$ with midpoint $r_5 \in D$, $r_4,r'_4 \in N_4$, $s_5 \in N_5$, $r'_3 \in N_3$. 

\medskip

Finally assume that $(r_3,r_4,r_5,r'_4,r'_3)$ induce a $P_5$ in $G$ with midpoint $r_5 \in D$, $r_4,r'_4 \in N_4$, $r_3,r'_3 \in N_3$, i.e., 
$r_3r'_4 \notin E$, $r'_3r_4 \notin E$. Then $r_3$ must have a neighbor $r_2 \in N_2$, i.e., $r_2r_3 \in E$. But then $(r_2,r_3,r_4,r_5,r'_4)$ induce a $P_5$ in $G$, which is a contradiction by Lemma~\ref{N6emptyN5indep}.

\medskip

Thus, $r_5 \in D \cap N_5$ is no $D$-midpoint of any $P_5$ in $G$, and Lemma \ref{N5noDmidpointP5} is shown.
\qed

\section{When no $D$-vertex is midpoint of a $P_5$ in $G$}\label{noDvertexmidpointP5}

In this section, every $D$-vertex is no midpoint of a $P_5$ in $G$.

\begin{lemma}\label{P8Pu1u2v2u3v3v4excluded}
For every $P_8$ $P=(u_1,v_1,u_2,v_2,u_3,v_3,u_4,v_4)$ in $G$, $u_1,u_2,v_2,u_3,v_3,v_4$ are excluded. 
\end{lemma}

\noindent
{\em Proof.}
Let $P=(u_1,v_1,u_2,v_2,u_3,v_3,u_4,v_4)$ induce a $P_8$ in $G$ and no $D$-vertex is midpoint of a $P_5$ in $G$. Recall that in this section,   
every $D$-vertex is no midpoint of a $P_5$ in $G$.

\medskip

Since $(u_1,v_1,u_2,v_2,u_3)$ induce a $P_5$ in $G$ and $u_2$ is midpoint of the $P_5$, $u_2 \notin D$. Analogously, 
since $(v_1,u_2,v_2,u_3,v_3)$ induce a $P_5$ in $G$ and $v_2$ is midpoint of the $P_5$, $v_2 \notin D$, 
since $(u_2,v_2,u_3,v_3,u_3)$ induce a $P_5$ in $G$ and $u_3$ is midpoint of the $P_5$, $u_3 \notin D$, and 
since $(v_2,u_3,v_3,u_4,v_4)$ induce a $P_5$ in $G$ and $v_3$ is midpoint of the $P_5$, $v_3 \notin D$, i.e., $u_2,v_2,u_3,v_3$ are excluded.

\medskip

Finally suppose to the contrary that either $u_1 \in D$ or $v_4 \in D$; without loss of generality, assume that $u_1 \in D$ (recall that $u_2,v_2$ are excluded). 
Then by the e.d.s.\ property, $v_1,u_2 \notin D$ and $u_2$ must have a $D$-neighbor $v \in D$ with $u_2v \in E$, $v \neq v_1,v_2$. 
Clearly, by the e.d.s.\ property, $u_1v \notin E$.  
Since in this section, $v$ is no midpoint of a $P_5$ in $G$, $vu_3 \notin E$ and $vu_4 \notin E$.
But then $(u_2,v,v_1,u_1,v_2,u_3,v_3,u_4,v_4)$ (with midpoint $u_2$) induce an $S_{1,2,5}$ in $G$, which is a contradiction. 
Thus, $u_1 \notin D$. Analogously, $v_4 \notin D$, i.e., $u_1,v_4$ are excluded, and Lemma \ref{P8Pu1u2v2u3v3v4excluded} is shown.
\qed

\begin{lemma}\label{DcapP8v1u4forced} 
For every $P_8$ $P=(u_1,v_1,u_2,v_2,u_3,v_3,u_4,v_4)$ in $G$, there are exactly two $D$-vertices $v_1 \in D$ and $u_4 \in D$ in $P$ which are forced.
\end{lemma}

\noindent
{\em Proof.}
Let $P=(u_1,v_1,u_2,v_2,u_3,v_3,u_4,v_4)$ induce a $P_8$ in $G$, and in this section, no $D$-vertex is midpoint of a $P_5$ in $G$. Recall that 
by Lemma \ref{P8Pu1u2v2u3v3v4excluded}, $u_1,u_2,v_2,u_3,v_3,v_4$ are excluded.

\medskip

Suppose to the contrary that either $v_1 \notin D$ or $u_4 \notin D$; without loss of generality, assume that $v_1 \notin D$.
By Lemma \ref{P8Pu1u2v2u3v3v4excluded}, $u_1,u_2 \notin D$ and $v_1$ must have a $D$-neighbor $u \in D$ with $uv_1 \in E$. 
Moreover, by the e.d.s.\ property, $u_1$ must have a $D$-neighbor $v \in D$ with $u_1v \in E$. Since in this section, $u,v$ are no midpoints of a $P_5$ in $G$, 
$uv_2 \notin E$, $uv_3 \notin E$, and $uv_4 \notin E$, as well as $vu_3 \notin E$ and $vu_4 \notin E$. 

If $vu_2 \notin E$ then $(v_1,u,u_1,v,u_2,v_2,u_3,v_3,u_4)$ (with midpoint $v_1$) induce an $S_{1,2,5}$ in $G$, which is a contradiction.   
Thus, $vu_2 \in E$ but then $(u_2,v,v_1,u,v_2,u_3,v_3,u_4,v_4)$ (with midpoint $u_2$) induce an $S_{1,2,5}$ in $G$, which is a contradiction. 
Thus, $v_1 \in D$ is forced.
Analogously, $u_4 \in D$ is forced, i.e., $v_1,u_4$ are forced, and Lemma \ref{DcapP8v1u4forced} is shown. 
\qed 

\medskip

Since in this section, there is no such $D$-vertex which is midpoint of a $P_5$ in $G$ then by Lemma \ref{DcapP8v1u4forced}, for every $P_8$ $P=(u_1,v_1,\ldots,u_4,v_4)$ in $G$ with forced $D$-vertices $v_1,u_4 \in D$, one can update $D_{basis}:= D_{basis} \cup \{v_1,u_4\}$ and redefine the distance levels $N_i$, $i \ge 1$, with respect to $D_{basis}$ as above. Finally, $G \setminus (N_0 \cup N_1)$ is $P_8$-free bipartite, and ED is solvable in polynomial time for $P_8$-free bipartite graphs \cite{BraMos2021/3}. 

\medskip

Thus, in this section, the e.d.s.\ problem can be solved in polynomial time. Now assume that there are $D$-vertices which are midpoints of a $P_5$ in $G$. 

\section{When $D$-vertices are midpoints of a $P_5$ in $G$}\label{DvertexmidpointP5}

Recall that by Lemma \ref{N6emptyN5indep}, $N_6=\emptyset$, $N_5$ is independent, every edge in $N_4$ does not contact $N_5$, and for every edge $r_3s_3$ in $N_3$, there is no such $P_4$ $(r_3,s_3,r_4,r_5)$ with $r_4 \in N_4$ and $r_5 \in N_4 \cup N_5$. 
Moreover, by Lemma \ref{N5noDmidpointP5}, there are no such $D$-midpoints in $D \cap N_5$ of a $P_5$ in $G$. 

\medskip

Recall that in Section \ref{noDvertexmidpointP5} and by Lemma \ref{DcapP8v1u4forced}, the e.d.s.\ problem can be solved in polynomial time.
Now assume that there is at least one $D$-vertex $r_0 \in D$ which is midpoint of a $P_5$ $(r_2,r_1,r_0,r'_1,r'_2)$ in $G$, 
say $(r_2,r_1,r_0,r'_1,r'_2)$ induce a $P_5$ in $G$ with $r_0 \in N_0$, $r_1,r'_1 \in N_1$, $r_2,r'_2 \in N_2$. 

Recall that by the e.d.s.\ property, $r_2,r'_2 \notin D$ and by (\ref{uinN2neighbinN3}), $r_2,r'_2 \in N_2$ must have $D$-neighbors $r_3,r'_3 \in D \cap N_3$, say $r_2r_3 \in E$ and $r'_2r'_3 \in E$. Since $G$ is chordal bipartite, i.e., there is no such $C_6$ in $G$, $(r_3,r_2,r_1,r_0,r'_1,r'_2,r'_3)$ induce a $P_7$ in $G$ with $r_3,r'_3 \in D \cap N_3$, $r_3 \neq r'_3$.  
Recall that by Lemma \ref{P5u1v1u2v2u3u2inD}, $r_3,r'_3 \in D$ are $r_0$-forced. Now assume that $r_0,r_3,r'_3 \in N_0$ and $r_1,r_2,r'_1,r'_2 \in N_1$.  

\medskip

If $r_2$ or $r'_2$ does not have any neighbor in $N_3$ then there is no such e.d.s.\ in $G$ with $D_{basis}$ (recall (\ref{uinN2neighbinN3}) and 
(\ref{uinN2twoneighbinN3})). 

\medskip

Now assume that $(u_1,v_1,u_2,v_2,u_3,v_3,u_4)$ induce a $P_7$ in $G$ with midpoint $v_2 \in N_0$, $u_1,u_4 \in N_0$, $v_1,u_2,u_3,v_3 \in N_1$. Then also every 
$(u_1,v_2,u_4)$-forced $D$-vertex is in $N_0$, and we have: 
\begin{equation}\label{N0N1connected}
G[N_0 \cup N_1] \mbox{ is connected}.
\end{equation}

\begin{lemma}\label{DmidpointP7v1v3cojoinN2} 
If $(u_1,v_1,u_2,v_2,u_3,v_3,u_4)$ induce a $P_7$ in $G$ with midpoint $v_2 \in N_0$, $u_1,u_4 \in N_0$, $v_1,u_2,u_3,v_3 \in N_1$, then 
$\{v_1,v_3\} \cojoin N_2$.
\end{lemma}

\noindent
{\em Proof.}
Let $(u_1,v_1,u_2,v_2,u_3,v_3,u_4)$ induce a $P_7$ in $G$ with midpoint $v_2 \in N_0$, $u_1,u_4 \in N_0$, $v_1,u_2,u_3,v_3 \in N_1$.

\medskip

Suppose to the contrary that without loss of generality, $r_2v_1 \in E$ with $r_2 \in N_2$. Since $(r_2,v_1,u_2,v_2,u_3,v_3)$ does not induce a $C_6$ in $G$, 
$r_2v_3 \notin E$. But then $r_2$ is an endpoint of a $P_6$ $(r_2,v_1,u_2,v_2,u_3,v_3)$ whose remaining vertices are in $N_0 \cup N_1$, 
which is a contradiction by Lemma \ref{uinN2endpointnoP6N0N1}. 
Thus, $r_2v_1 \notin E$, i.e., $\{v_1\} \cojoin N_2$, and analogously, $\{v_3\} \cojoin N_2$,  i.e., $\{v_1,v_3\} \cojoin N_2$, and 
Lemma \ref{DmidpointP7v1v3cojoinN2} is shown. 
\qed

\begin{lemma}\label{DmidpointP7N(u1)N(u4)cojoinN2} 
If $(u_1,v_1,u_2,v_2,u_3,v_3,u_4)$ induce a $P_7$ in $G$ with midpoint $v_2 \in N_0$, $u_1,u_4 \in N_0$, $v_1,u_2,u_3,v_3 \in N_1$, and $dist_G(u_1,u_4)=6$
then $N(u_1) \cojoin N_2$ and $N(u_4) \cojoin N_2$.
\end{lemma}

\noindent
{\em Proof.}
Let $(u_1,v_1,u_2,v_2,u_3,v_3,u_4)$ induce a $P_7$ in $G$ with midpoint $v_2 \in N_0$, $u_1,u_4 \in N_0$, $v_1,u_2,u_3,v_3 \in N_1$.
Without loss of generality, $u_i \in X$, $1 \le i \le 4$ and $v_j \in Y$, $1 \le j \le 3$.  

\medskip

Suppose to the contrary that $r_2 \in N_2$ contacts $N(u_1)$ or $N(u_4)$, say $r_1r_2 \in E$, $r_1 \in N_1$, with $u_1r_1 \in E$ or $u_4r_1 \in E$; 
without loss of generality, $u_1r_1 \in E$. Recall that $r_2v_1 \notin E$ and $r_2v_3 \notin E$. 

\medskip

 Since $dist_G(u_1,u_4)=6$, $r_1u_3 \notin E$. Recall that by Lemma \ref{uinN2endpointnoP6N0N1}, 
$r_2$ is no endpoint of a $P_6$ whose remaining vertices are in $N_0 \cup N_1$.
Thus, $r_1u_2 \in E$. But then again $r_2$ is an endpoint of a $P_6$ $(r_2,r_1,u_2,v_2,u_3,v_3)$ whose remaining vertices are in $N_0 \cup N_1$, 
which is a contradiction. Thus, $N(u_1) \cojoin N_2$. Analogously, $N(u_4) \cojoin N_2$, and Lemma~\ref{DmidpointP7N(u1)N(u4)cojoinN2} is shown.
\qed

\begin{lemma}\label{DmidpointP7distu1u4=4} 
If $(u_1,v_1,u_2,v_2,u_3,v_3,u_4)$ induce a $P_7$ in $G$ with midpoint $v_2 \in N_0$, $u_1,u_4 \in N_0$, $v_1,u_2,u_3,v_3 \in N_1$, then $dist_G(u_1,u_4)=4$.
\end{lemma}

\noindent
{\em Proof.}
Let $(u_1,v_1,u_2,v_2,u_3,v_3,u_4)$ induce a $P_7$ in $G$ with midpoint $v_2 \in N_0$, $u_1,u_4 \in N_0$, $v_1,u_2,u_3,v_3 \in N_1$.
Without loss of generality, $u_i \in X$, $1 \le i \le 4$ and $v_j \in Y$, $1 \le j \le 3$.  

\medskip

Suppose to the contrary that $dist_G(u_1,u_4) \neq 4$, i.e., $dist_G(u_1,u_4)=6$. Then by Lemma \ref{DmidpointP7N(u1)N(u4)cojoinN2}, $N(u_1) \cojoin N_2$ and $N(u_4) \cojoin N_2$. 
Then for every $r_2 \in N_2 \cap X$, $r_2 \cojoin N(u_1) \cup N(u_4)$, say $r_1r_2 \in E$ with $r_1 \in N_1$, $r_1u_1 \notin E$, $r_1u_4 \notin E$, and $r_0r_1 \in E$ with $r_0 \in N_0$, $r_0 \neq u_1,u_4$. By (\ref{uinN2neighbinN3}), $r_2$ must have a $D$-neighbor in $N_3$, and by (\ref{uinN2twoneighbinN3}), $|N(r_2)| \ge 2$, say $r_3,r'_3 \in N(r_2) \cap N_3$, and without loss of generality, $r_3 \in D$. Then by the e.d.s.\ property, $r'_3 \notin D$ and $r'_3$ must have a $D$-neighbor $r'_4 \in D \cap (N_3 \cup N_4)$.

\medskip

First assume that $r_1$ contacts $N(v_2)$. 

\medskip

If $r_1u_2 \in E$ and $r_1u_3 \in E$ then $(r_1,r_2,r'_3,r'_4,u_2,v_1,u_1,u_3,v_3,u_4)$ (with midpoint $r_1$)
induce an $S_{3,3,3}$ in $G$, which is a contradiction. Thus, $r_1u_2 \notin E$ or $r_1u_3 \notin E$, say without loss of generality, $r_1u_2 \notin E$.
Since by Lemma \ref{uinN2endpointnoP6N0N1}, $r_2$ is no endpoint of a $P_6$ $(r_2,r_1,u_3,v_2,u_2,v_1)$ whose remaining vertices are in $N_0 \cup N_1$, 
$r_1u_3 \notin E$.

If $r_1u \in E$, $u \neq u_2,u_3$, with $v_2u \in E$ then, since $r_2$ is no endpoint of a $P_6$ $(r_2,r_1,u,v_2,u_2,v_1)$ whose remaining vertices are in 
$N_0 \cup N_1$, $uv_1 \in E$. 
Moreover, since $r_2$ is no endpoint of a $P_6$ $(r_2,r_1,u,v_2,u_3,v_3)$ whose remaining vertices are in $N_0 \cup N_1$, $uv_3 \in E$. 
But then $(u_1,v_1,u,v_3,u_4)$ induce a $P_5$ in $G[N_0 \cup N_1]$, i.e., $dist_G(u_1,u_4)=4$.

\medskip

Next assume that $r_1$ does not contact $N(v_2)$. Without loss of generality, assume that $dist_G(r_0,v_2)=3$, say $(r_0,r'_1,u,v_2)$ induce a $P_4$ in 
$G[N_0 \cup N_1]$ (possibly $u=u_2$ or $u=u_3$).
Since by Lemma \ref{uinN2endpointnoP6N0N1}, $r_2$ is no endpoint of a $P_6$ $(r_2,r_1,r_0,r'_1,u,v_2)$ whose remaining vertices are in $N_0 \cup N_1$, 
$r'_1r_2 \in E$.
If $r'_1u_2 \in E$ and $r'_1u_3 \in E$ then $(r'_1,u_2,v_1,u_1,u_3,v_3,u_4,r_2,r'_3,r'_4)$ (with midpoint $r'_1$) induce an $S_{3,3,3}$ in $G$, which is a contradiction. Thus, $r'_1u_2 \notin E$ or $r'_2u_3 \notin E$; without loss of generality, $r'_1u_2 \notin E$.

Since by Lemma \ref{uinN2endpointnoP6N0N1}, $r_2$ is no endpoint of a $P_6$ $(r_2,r'_1,u,v_2,u_2,v_1)$ whose remaining vertices are in $N_0 \cup N_1$, 
$uv_1 \in E$. If $u=u_3$ then $dist_G(u_1,u_4)=4$. Now assume that $u \neq u_3$. 
Since by Lemma \ref{uinN2endpointnoP6N0N1}, $r_2$ is no endpoint of a $P_6$ $(r_2,r'_1,u,v_2,u_3,v_3)$ whose remaining vertices are in $N_0 \cup N_1$, 
$uv_3 \in E$. But then again $dist_G(u_1,u_4)=4$, and Lemma \ref{DmidpointP7distu1u4=4} is shown.
\qed 

\section{When $(u_1,v_1,u_2,v_2,u_3,v_3,u_4)$ is a $P_7$ in $G[N_0 \cup N_1]$ with midpoint $v_2 \in N_0 \cap Y$ and $u_1,u_4 \in N_0 \cap X$}\label{P7N0N1u1u4inX}

\begin{lemma}\label{DYN0midpointP7} 
If $(u_1,v_1,u_2,v_2,u_3,v_3,u_4)$ induce a $P_7$ in $G$ with midpoint $v_2 \in N_0 \cap Y$, $u_1,u_4 \in N_0 \cap X$, $v_1,u_2,u_3,v_3 \in N_1$   
then $N_5 \cap Y=\emptyset$ and $N_4$ is independent. 
\end{lemma}

\noindent
{\em Proof.}
Let $(u_1,v_1,u_2,v_2,u_3,v_3,u_4)$ induce a $P_7$ in $G$ with midpoint $v_2 \in N_0 \cap Y$, $u_1,u_4 \in N_0 \cap X$, $v_1,u_2,u_3,v_3 \in N_1$.  

\medskip

Suppose to the contrary that $N_5 \cap Y \neq \emptyset$, i.e., $(r_2,r_3,r_4,r_5)$ induce a $P_4$ in $G$ with $r_5 \in N_5 \cap Y$, $r_4 \in N_4 \cap X$,  
$r_3 \in N_3 \cap Y$, $r_2 \in N_2 \cap X$. Since $N(v_2) \subset X$, $r_2 \cojoin N(v_2)$. 
By Lemma \ref{DmidpointP7v1v3cojoinN2}, $r_2v_1 \notin E$ and $r_2v_3 \notin E$, say $r_1r_2 \in E$ with 
$r_1 \in N_1 \cap Y$, $r_1 \neq v_1,v_3$. 

\medskip

First assume that $r_1u_1 \in E$ or $r_1u_4 \in E$; without loss of generality, $r_1u_1 \in E$. 
Since by Lemma \ref{uinN2endpointnoP6N0N1}, $r_2$ is no endpoint of a $P_6$ $(r_2,r_1,u_1,v_1,u_2,v_2)$ whose remaining vertices are in $N_0 \cup N_1$, 
$r_1u_2 \in E$. Moreover, since by Lemma \ref{uinN2endpointnoP6N0N1}, $r_2$ is no endpoint of a $P_6$ $(r_2,r_1,u_2,v_2,u_3,v_3)$ whose remaining vertices are in $N_0 \cup N_1$, $r_1u_3 \in E$.
But then $(u_3,v_2,v_3,u_4,r_1,r_2,r_3,r_4,r_5)$ (with midpoint $u_3$) induce an $S_{1,2,5}$ in $G$, which is a contradiction. Thus, $r_1u_1 \notin E$, 
and analogously, $r_1u_4 \notin E$, i.e., $r_0r_1 \in E$ with $r_0 \in N_0$, $r_0 \neq u_1,u_4$.

\medskip
 
Since by (\ref{N0N1connected}), $G[N_0 \cup N_1]$ is connected, $dist_G(r_0,v_2) \ge 3$; without loss of generality, assume that $dist_G(r_0,v_2)=3$.

Since $(u_2,v_2,v_1,u_1,r_1,r_2,r_3,r_4,r_5)$ (with midpoint $u_2$) does not induce an $S_{1,2,5}$ in $G$, $r_1u_2 \notin E$.
Analogously, since $(u_3,v_2,v_3,u_4,r_1,r_2,r_3,r_4,r_5)$ (with midpoint $u_3$) does not induce an $S_{1,2,5}$ in $G$, $r_1u_3 \notin E$. 

If $r_1u \in E$ with $u \in N(v_2)$ then since $(v_2,u_2,v_1,u_1,u_3,v_3,u_4,u,r_1,r_2)$ (with midpoint $v_2$) does not induce an $S_{3,3,3}$ in $G$, 
$uv_1 \in E$ or $uv_3 \in E$; without loss of generality, $uv_1 \in E$. 
But then $(u,v_2,v_1,u_1,r_1,r_2,r_3,r_4,r_5)$ (with midpoint $u$) induce an $S_{1,2,5}$ in $G$, which is a contradiction. Thus, $r_1 \cojoin N(v_2)$.   

Let $(r_0,r'_1,u,v_2)$ induce a $P_4$ in $G$ with $r_0r'_1 \in E$, $r'_1 \neq r_1$, and $v_2u \in E$ (possibly $u=u_2$ or $u=u_3$). 
Since by Lemma \ref{uinN2endpointnoP6N0N1}, $r_2$ is no endpoint of a $P_6$ $(r_2,r_1,r_0,r'_1,u,v_2)$ whose remaining vertices are in $N_0 \cup N_1$, 
$r'_1r_2 \in E$. 

If $u=u_2$ then $(u_2,v_2,v_1,u_1,r'_1,r_2,r_3,r_4,r_5)$ (with midpoint $u_2$) induce an $S_{1,2,5}$ in $G$, which is a contradiction. 
Thus, $r'_1u_2 \notin E$ and analogously, $r'_1u_3 \notin E$, say $r'_1u \in E$ with $u \in N(v_2)$, $u \neq u_2,u_3$. 

Since $(v_2,u_2,v_1,u_1,u_3,v_3,u_4,u,r'_1,r_2)$ (with midpoint $v_2$) does not induce an $S_{3,3,3}$ in $G$, $uv_1 \in E$ or $uv_3 \in E$; without loss of generality, $uv_1 \in E$. But then $(u,v_2,v_1,u_1,r'_1,r_2,r_3,r_4,r_5)$ (with midpoint $u_2$) induce an $S_{1,2,5}$ in $G$, which is a contradiction.  

\medskip 

Thus, $N_5 \cap Y=\emptyset$. Analogously, $N_4$ is independent (else there are $S_{1,2,5}$'s or $S_{3,3,3}$'s in $G$), and Lemma \ref{DYN0midpointP7} is shown.
\qed     

\begin{coro}\label{coroDtwomidpointsP7} 
If $(u_1,v_1,u_2,v_2,u_3,v_3,u_4)$ induce a $P_7$ in $G$ with midpoint $v_2 \in N_0 \cap Y$, $u_1,u_4 \in N_0 \cap X$, 
and $(v_4,u_5,v_5,u_6,v_6,u_7,v_7)$ induce a $P_7$ in $G$ with midpoint $u_6 \in N_0 \cap X$, $v_4,v_7 \in N_0 \cap Y$,   
then $N_5=\emptyset$. 
\end{coro}

Let $r_4 \in D \cap N_4$ be a midpoint of a $P_5$ in $N_2 \cup N_3 \cup N_4 \cup N_5$.
Recall that by Lemma~\ref{N6emptyN5indep}, every $r_2 \in N_2$ is no such endpoint of a $P_5$ whose remaining vertices are in $N_3 \cup N_4 \cup N_5$. 

\begin{lemma}\label{noDN4XmidpointP5} 
There are no such $D$-vertices in $D \cap N_4 \cap X$ with midpoint $P_5$. 
\end{lemma}

\noindent
{\em Proof.}
Recall that $(u_1,v_1,u_2,v_2,u_3,v_3,u_4)$ induce a $P_7$ in $N_0 \cup N_1$ with midpoint $v_2 \in N_0 \cap Y$, $u_1,u_4 \in N_0 \cap X$ and 
$v_1,u_2,u_3,v_3 \in N_1$.

\medskip

Suppose to the contrary that there is a $D$-vertex $r_4 \in D \cap N_4 \cap X$ with midpoint of a $P_5$. Recall that by Lemma \ref{DYN0midpointP7}, 
 $N_5 \cap Y=\emptyset$ and $N_4$ is independent. Thus, $r_4 \cojoin N_4 \cup N_5$.  

\medskip

\noindent
{\bf Case 1.} $r_4$ is midpoint of a $P_5$ $(r_2,r_3,r_4,r'_3,r'_2)$ in $G$ with $r_2,r'_2 \in N_2$ and $r_3,r'_3 \in N_3$:

\medskip

Then $r_1r_2 \in E$ with $r_1 \in N_1$ and $r_0r_1 \in E$ with $r_0 \in N_0$ (possibly $r_0=u_1$ or $r_0=u_4$). Since $(r_1,r_2,r_3,r_4,r'_3,r'_2)$ does not induce a $C_6$ in $G$, $r_1r'_2 \notin E$. Moreover, $r_2,r'_2$ must have $D$-neighbors in $D \cap N_3$, say $r_2s_3 \in E$ with $s_3 \in D \cap N_3$, and since $(s_3,r_2,r_3,r_4,r'_3,r'_2)$ does not induce a $C_6$ in $G$, 
$s_3r'_2 \notin E$, i.e., $s'_3r'_2 \in E$ with $s'_3 \in D \cap N_3$, $s_3 \neq s'_3$. 
But then $(r_2,s_3,r_1,r_0,r_3,r_4,r'_3,r'_2,s'_3)$ (with midpoint $r_2$) induce an $S_{1,2,5}$ in $G$, which is a contradiction.  

Thus, $r_4$ is no such midpoint of a $P_5$ $(r_2,r_3,r_4,r'_3,r'_2)$ in $G$ with $r_2,r'_2 \in N_2$ and $r_3,r'_3 \in N_3$.

\medskip

\noindent
{\bf Case 2.} $r_4$ is midpoint of a $P_5$ $(r_2,r_3,r_4,r'_3,s'_3)$ in $G$ with $r_2 \in N_2$ and $r_3,r'_3,s'_3 \in N_3$:

\medskip

Then $r_2$ is the endpoint of a $P_5$ $(r_2,r_3,r_4,r'_3,s'_3)$, which is a contradiction by Lemma~\ref{N6emptyN5indep}. 
Thus, $r_4$ is no such midpoint of a $P_5$ $(r_2,r_3,r_4,r'_3,s'_3)$ in $G$ with $r_2 \in N_2$ and $r_3,r'_3,s'_3 \in N_3$.

\medskip

\noindent
{\bf Case 3.} $r_4$ is midpoint of a $P_5$ $(s_3,r_3,r_4,r'_3,s'_3)$ in $G$ with $s_3,r_3,r'_3,s'_3 \in N_3$:

\medskip

Let $s_2 \in N_2$ with $s_2s_3 \in E$. But then $s_2$ is the endpoint of a $P_5$ $(s_2,s_3,r_3,r_4,r'_3)$, which is a contradiction by Lemma~\ref{N6emptyN5indep}. 
Thus, $r_4$ is no such midpoint of a $P_5$ $(s_3,r_3,r_4,r'_3,s'_3)$ in $G$ with $s_3,r_3,r'_3,s'_3 \in N_3$.

\medskip

\noindent
{\bf Case 4.} $r_4$ is midpoint of a $P_5$ $(s_3,r_3,r_4,r'_3,r'_4)$ in $G$ with $s_3,r_3,r'_3 \in N_3$ and $r'_4 \in N_4$:

\medskip

Let $s_2 \in N_2$ with $s_2s_3 \in E$. But then $s_2$ is the endpoint of a $P_5$ $(s_2,s_3,r_3,r_4,r'_3)$, which is a contradiction by Lemma~\ref{N6emptyN5indep}. 
Thus, $r_4$ is no such midpoint of a $P_5$ $(s_3,r_3,r_4,r'_3,r'_4)$ in $G$ with $s_3,r_3,r'_3 \in N_3$ and $r'_4 \in N_4$.

\medskip

\noindent
{\bf Case 5.} $r_4$ is midpoint of a $P_5$ $(s_4,r_3,r_4,r'_3,s'_4)$ in $G$ with $r_3,r'_3 \in N_3$ and $s_4,s'_4 \in N_4$:

\medskip

Let $r_2 \in N_2 \cap X$ with $r_2r_3 \in E$. By Lemma~\ref{N6emptyN5indep}, $r_2$ is no such endpoint of a $P_5$ $(r_2,r_3,r_4,r'_3,s'_4)$, i.e., 
$r_2r'_3 \in E$. Recall that by Lemma \ref{DmidpointP7v1v3cojoinN2}, $r_2v_1 \notin E$, $r_2v_3 \notin E$, say $r_1r_2 \in E$, $r_1 \neq v_1,v_3$, and 
since by Lemma \ref{DYN0midpointP7}, $N_5 \cap Y=\emptyset$ and $N_4$ is independent, 
$s_4,s'_4$ must have $D$-neighbors in $N_3 \cap Y$, say $t_3 \in D \cap N_3 \cap Y$ with $t_3s_4 \in E$ and $t'_3 \in D \cap N_3 \cap Y$ with $t'_3s'_4 \in E$. 
Since there is no such $C_6$ in $G$, $t_3 \neq t'_3$. 

By the e.d.s.\ property, $r_2t_3 \notin E$ or $r_2t'_3 \notin E$; without loss of generality, $r_2t'_3 \notin E$. 
If $u_1r_1 \in E$ then, since by Lemma \ref{uinN2endpointnoP6N0N1}, $r_2$ is no such endpoint of a $P_6$ whose remaining vertices are in $N_0 \cup N_1$, 
$r_1u_2 \in E$ and analogously, $r_1u_3 \in E$. 
But then $(u_3,v_2,v_3,u_4,r_1,r_2,r'_3,s'_4,t'_3)$ (with midpoint $u_3$) induce an $S_{1,2,5}$ in $G$, which is a contradiction. 
Thus, $u_1r_1 \notin E$ and analogously, $u_4r_1 \notin E$, say $r_0r_1 \in E$ with $r_0 \in N_0 \cap X$, $r_0 \neq u_1,u_4$. 

Without loss of generality, $dist_G(r_0,v_2)=3$, say $(r_0,r'_1,u,v_2)$ induce a $P_4$ in $G[N_0 \cup N_1]$ (possibly $u=u_2$ or $u=u_3$). 
Recall that $r_1u_2 \notin E$ and $r_1u_3 \notin E$. Since by Lemma \ref{uinN2endpointnoP6N0N1}, $r_2$ is no endpoint of a $P_6$ $(r_2,r_1,r_0,r'_1,u,v_2)$, 
$r'_1r_2 \in E$. 

If $u=u_2$ then $(u_2,v_2,v_1,u_1,r'_1,r_2,r'_3,s'_4,t'_3)$ (with midpoint $u_2$) induce an $S_{1,2,5}$ in $G$, which is a contradiction. 
Thus, $u \neq u_2$. 

Moreover, if $u=u_3$ then $(u_3,v_2,v_3,u_4,r'_1,r_2,r'_3,s'_4,t'_3)$ (with midpoint $u_3$) induce an $S_{1,2,5}$ in $G$, which is a contradiction. 
Thus, $u \neq u_3$. 

Moreover, since by Lemma \ref{uinN2endpointnoP6N0N1}, $r_2$ is no endpoint of a $P_6$ $(r_2,r'_1,u,v_2,u_3,v_2)$, $r'_1u_3 \in E$. 
But then $(u_3,v_2,v_3,u_4,r'_1,r_2,r'_3,s'_4,t'_3)$ (with midpoint $u_3$) induce an $S_{1,2,5}$ in $G$, which is a contradiction. 

\medskip

Thus, there is no such $r_4 \in D \cap N_4 \cap X$ with midpoint of a $P_5$ in $G$, and Lemma~\ref{noDN4XmidpointP5} is shown.
\qed 

\begin{coro}\label{corotwoP7inN0N1midpointsXYinN0noDN4midptP5}
If $(u_1,v_1,u_2,v_2,u_3,v_3,u_4)$ induce a $P_7$ in $N_0 \cup N_1$ with midpoint $v_2 \in N_0 \cap Y$, $u_1,u_4 \in N_0 \cap X$ and 
$(v_4,u_4,v_5,u_5,v_6,u_6,v_7)$ induce a $P_7$ in $N_0 \cup N_1$ with midpoint $u_5 \in N_0 \cap X$, $v_4,v_7 \in N_0 \cap Y$ then there are no such 
 $D$-vertices in $N_4$ with midpoints of a $P_5$ in $G$. 
\end{coro}

\begin{lemma}\label{atmostoneDN3Y} 
There is at most one $D$-vertex $r_3 \in D \cap N_3 \cap Y$. 
\end{lemma}

\noindent
{\em Proof.}
Recall that $P=(u_1,v_1,u_2,v_2,u_3,v_3,u_4)$ induce a $P_7$ in $G[N_0 \cup N_1]$ with midpoint $v_2 \in N_0 \cap Y$, $u_1,u_4 \in N_0 \cap X$, $v_1,u_2,u_3,v_3 \in N_1$, and by (\ref{vinN3neighbinN3N4}), for $r_3 \in D \cap N_3 \cap Y$, $(r_2,r_3,r_4)$ induce a $P_3$ in $G$ with $r_2 \in N_2$, $r_4 \in N_3 \cup N_4$. 

\medskip

Suppose to the contrary that there are two $D$-vertices $r_3,s_3 \in D \cap N_3 \cap Y$, i.e., $(r_2,r_3,r_4)$, $(s_2,s_3,s_4)$ with $s_2 \in N_2$, 
$s_4 \in N_3 \cup N_4$, induce a $2P_3$ in $G$ with $r_2,s_2 \in N_2 \cap X$, $r_4,s_4 \in (N_3 \cup N_4) \cap X$.

\medskip

Recall that by Lemma \ref{DmidpointP7v1v3cojoinN2}, $r_2v_i \notin E$, $s_2v_i \notin E$, $i \in \{1,3\}$.
Moreover, recall that by Lemma \ref{2P3N3commonN2neighbor}, for $2P_3$ $(r_2,r_3,r_4)$, $(s_2,s_3,s_4)$ in $G$ with $r_3,s_3 \in D \cap N_3 \cap Y$, $r_2$ and $s_2$ must have a common neighbor $r_1 \in N_1$ with $r_1r_2 \in E$ and $r_1s_2 \in E$. Moreover $r_0r_1 \in E$ with $r_0 \in N_0$ (possibly $r_0=u_1$ or $r_0=u_4$). 

\medskip

First assume that $r_1u_1 \in E$ or $r_1u_4 \in E$; without loss of generality, $r_1u_1 \in E$ and by the e.d.s.\ property, $r_1u_4 \notin E$.  
Since by Lemma \ref{uinN2endpointnoP6N0N1}, $r_2$ is no endpoint of a $P_6$ $(r_2,r_1,u_1,v_1,u_2,v_2)$ whose remaining vertices are in $N_0 \cup N_1$, 
$r_1u_2 \in E$. 
Moreover, since by Lemma \ref{uinN2endpointnoP6N0N1}, $r_2$ is no endpoint of a $P_6$ $(r_2,r_1,u_2,v_2,u_3,v_3)$ whose remaining vertices are in $N_0 \cup N_1$, $r_1u_3 \in E$.
But then $(r_1,r_2,r_3,r_4,s_2,s_3,s_4,u_3,v_3,u_4)$ (with midpoint $r_1$) induce an $S_{3,3,3}$ in $G$, which is a contradiction. 
Thus, $r_1u_1 \notin E$ and analogously, $r_1u_4 \notin E$, say $r_0r_1 \in E$ with $r_0 \in N_0$, $r_0 \neq u_1,u_4$. 

\medskip

Since $(r_1,r_2,r_3,r_4,s_2,s_3,s_4,u_2,v_1,u_1)$ (with midpoint $r_1$) does not induce an $S_{3,3,3}$ in $G$, $r_1u_2 \notin E$. 
Analogously, since $(r_1,r_2,r_3,r_4,s_2,s_3,s_4,u_3,v_3,u_4)$ (with midpoint $r_1$) does not induce an $S_{3,3,3}$ in $G$, $r_1u_3 \notin E$. 
Moreover, if $r_1u \in E$ with $u \in N(v_2)$, $u \neq u_2,u_3$, then $(r_1,r_2,r_3,r_4,s_2,s_3,s_4,u,v_2,u_3)$ (with midpoint $r_1$) induce an $S_{3,3,3}$ in $G$, 
which is a contradiction. Thus, $r_1 \cojoin N(v_2)$.

Without loss of generality, assume that $dist_G(r_0,v_2)=3$, say $(r_0,r'_1,u,v_2)$ induce a $P_4$ in $G[N_0 \cup N_1]$ (possibly $u=u_2$ or $u=u_3$), 
i.e., $r'_1 \neq r_1$. 

Since by Lemma \ref{uinN2endpointnoP6N0N1}, $r_2$ is no endpoint of a $P_6$ $(r_2,r_1,r_0,r'_1,u,v_2)$, $r'_1r_2 \in E$, and analogously, 
since by Lemma \ref{uinN2endpointnoP6N0N1}, $s_2$ is no endpoint of a $P_6$ $(s_2,r_1,r_0,r'_1,u,v_2)$, $r'_1s_2 \in E$. 

If $r'_1u_2 \in E$ then $(r'_1,r_2,r_3,r_4,s_2,s_3,s_4,u_2,v_1,u_1)$ (with midpoint $r'_1$) induce an $S_{3,3,3}$ in $G$, which is a contradiction. 
Thus, $r'_1u_2 \notin E$, and analogously, $r'_1u_3 \notin E$, i.e., $r'_1u \in E$ with $u \in N(v_2)$, $u \neq u_2,u_3$. 
But then $(r'_1,r_2,r_3,r_4,s_2,s_3,s_4,u,v_2,u_2)$ (with midpoint $r'_1$) induce an $S_{3,3,3}$ in $G$, which is a contradiction.   

\medskip

Thus, there is at most one $D$-vertex $r_3 \in D \cap N_3 \cap Y$, and Lemma \ref{atmostoneDN3Y} is shown.
\qed

\begin{coro}\label{coroDN3XorY} 
There are at most two such $D$-vertices in $N_3 \cap X$ and at most one such $D$-vertex in $N_3 \cap Y$. 
\end{coro}

Recall that $P=(u_1,v_1,u_2,v_2,u_3,v_3,u_4)$ induce a $P_7$ in $N_0 \cup N_1$ with midpoint $v_2 \in N_0 \cap Y$, $u_1,u_4 \in N_0 \cap X$ and 
$v_1,u_2,u_3,v_3 \in N_1$.

\medskip

If $N_4=\emptyset$ then the e.d.s.\ property can be done in polynomial time. Now assume that $N_4 \neq \emptyset$. 

\medskip

By Lemmas \ref{atmosttwoDN3XorY}, \ref{atmostoneDN3Y} and Corollary \ref{coroDN3XorY}, for at most two such $D$-vertices $r_3,s_3 \in D \cap N_3 \cap X$ and at most one such $D$-vertex $t_3 \in D \cap N_3 \cap Y$, one can update $D_{basis}:= D_{basis} \cup \{r_3,s_3,t_3\}$ and redefine the distance levels $N_i$, $i \ge 1$, with respect to $D_{basis}$ as above, i.e., $N_4 \cap (N(r_3) \cup N(s_3) \cup N(t_3)) \subset N'_1$.

Since $N_2$ contacts $t_3 \in D \cap N_3 \cap Y$ or $r_3,s_3 \in D \cap N_3 \cap X$, $N_2 \subset N'_1$, and $N_3 \setminus D \subset N'_2$. 
Then $N'_5 =\emptyset$. Moreover, $N_4 \setminus (N(r_3) \cup N(s_3) \cup N(t_3)) \subset N'_3$. 

\subsection{When there are $P_7$'s in $G$ with midpoints in $D \cap X$ and $D \cap Y$}\label{2P7smidpointsDXandDY}

Let $(u_1,v_1,u_2,v_2,u_3,v_3,u_4)$ induce a $P_7$ in $N_0 \cup N_1$ with midpoint $v_2 \in N_0 \cap Y$, $u_1,u_4 \in N_0 \cap X$ and 
$(v_4,u_4,v_5,u_5,v_6,u_6,v_7)$ induce a $P_7$ in $N_0 \cup N_1$ with midpoint $u_5 \in N_0 \cap X$, $v_4,v_7 \in N_0 \cap Y$.

Recall that by Lemma \ref{N5noDmidpointP5}, there are no $D$-midpoints in $D \cap N_5$ of a $P_5$ in $G$.
Then by Corollary \ref{corotwoP7inN0N1midpointsXYinN0noDN4midptP5}, there are no such $D$-vertices in $N_4$ with midpoints of a $P_5$ in $G$. 
Recall that by Lemma \ref{atmostoneDN3Y}, there is at most one $D$-vertex $r_3 \in D \cap N_3 \cap Y$, and analogously, there is at most one $D$-vertex 
$s_3 \in D \cap N_3 \cap X$. Then one can update $D_{basis}:= D_{basis} \cup \{r_3,s_3\}$ and redefine the distance levels $N_i$, $i \ge 1$, with respect to $D_{basis}$ as above, i.e., $N_4 \cap (N(r_3) \cup N(s_3)) \subset N'_1$. Moreover, $N_4 \setminus (N(r_3) \cup N(s_3)) \subset N'_3$. 
Then the e.d.s.\ problem can be solved in polynomial time 

\subsection{When there are no $P_7$'s in $G$ with either midpoints in $D \cap X$ or midpoints in $D \cap Y$}\label{noP7midpointDXorDY}

Without loss of generality, there is no such $P_7$ in $G$ with midpoint in $D \cap X$. Now assume that there are $P_7$'s in $G$ with midpoints in $D \cap Y$. 
If there is only one such $P_7$ with midpoints in $D \cap Y$ then by Section \ref{noDvertexmidpointP5}, the e.d.s.\ problem can be solved in polynomial time. 
Now assume that there are at least two such $P_7$'s with midpoints in $D \cap Y$, say $P=(u_1,v_1,u_2,v_2,u_3,v_3,u_4)$, $P'=(u'_1,v'_1,u'_2,v'_2,u'_3,v'_3,u'_4)$ induce $P_7$'s in $G$ with midpoints $v_2,v'_2 \in D \cap Y$, $u_1,u'_1,u_4,u'_4 \in D \cap X$.

\begin{lemma}\label{P7PP'nodist1} 
For $P_7$'s $P=(u_1,v_1,u_2,v_2,u_3,v_3,u_4)$, $P'=(u'_1,v'_1,u'_2,v'_2,u'_3,v'_3,u'_4)$ with midpoints $v_2,v'_2 \in D \cap Y$, $u_1,u'_1,u_4,u'_4 \in D \cap X$, 
every vertex in $P$ does not contact $P'$, i.e., $dist_G(P,P') \ge 2$.
\end{lemma}

\noindent
{\em Proof.}
Let $P=(u_1,v_1,u_2,v_2,u_3,v_3,u_4)$, $P'=(u'_1,v'_1,u'_2,v'_2,u'_3,v'_3,u'_4)$ induce $P_7$'s with midpoints $v_2,v'_2 \in D \cap Y$ and 
$u_1,u'_1,u_4,u'_4 \in D \cap X$. By the e.d.s.\ property, $u_1,v_2,u_4$ do not contact $P'$ and $u'_1,v'_2,u'_4$ do not contact $P$.

\medskip

Suppose to the contrary that a vertex $v_1,u_2,v_2,u_3 \in P$ contacts $P'$.  

\medskip

Without loss of generality, $u_2v'_1 \in E$. By the e.d.s.\ property, $u_2v'_2 \notin E$, and since $(u_2,v'_1,u'_2,v'_2,u'_3,v'_3)$ does not induce a $C_6$ in $G$, $u_2v'_3 \notin E$.

But then $(v'_1,u'_1,u_2,v_2,u'_2,v'_2,u'_3,v'_3,u'_4)$ (with midpoint $v'_1$) induce an $S_{1,2,5}$ in $G$, which is a contradiction. 
Thus, $u_2v'_1 \notin E$, and analogously, $u_2v'_3 \notin E$, $u_3v'_1 \notin E$, $u_3v'_3 \notin E$,
$u'_2v_1 \notin E$, $u'_2v_3 \notin E$, $u'_3v_1 \notin E$, $u'_3v_3 \notin E$.  

\medskip

Then $dist_G(P,P') \ge 2$ and Lemma \ref{P7PP'nodist1} is shown.
\qed

\begin{lemma}\label{P7PP'nodist2} 
For $P_7$'s $P=(u_1,v_1,u_2,v_2,u_3,v_3,u_4)$, $P'=(u'_1,v'_1,u'_2,v'_2,u'_3,v'_3,u'_4)$ with midpoints $v_2,v'_2 \in D \cap Y$, $u_1,u'_1,u_4,u'_4 \in D \cap X$, 
$dist_G(P,P') \ge 3$.
\end{lemma}

\noindent
{\em Proof.}
Let $P=(u_1,v_1,u_2,v_2,u_3,v_3,u_4)$, $P'=(u'_1,v'_1,u'_2,v'_2,u'_3,v'_3,u'_4)$ induce $P_7$'s with midpoints $v_2,v'_2 \in D \cap Y$ and 
$u_1,u'_1,u_4,u'_4 \in D \cap X$. Recall that by Lemma~\ref{P7PP'nodist1}, $dist_G(P,P') \ge 2$. 

\medskip

Suppose to the contrary that $dist_G(P,P')=2$. By the e.d.s.\ property, there is no common neighbor in $G \setminus (V(P) \cup V(P'))$ between some vertices in 
$u_1,v_2,u_4$ and $u'_1,v'_2,u'_4$. 

\medskip

First assume that without loss of generality, $v_1u \in E$ and $v'_1u \in E$ with $u \notin V(P) \cup V(P')$. 
Since $(v_1,u_1,u,v'_1,u_2,v_2,u_3,v_3,u_4)$ (with midpoint $v_1$) does not induce an $S_{1,2,5}$ in $G$, $uv_2 \in E$ or $uv_3 \in E$.
Since $(u,v_1,v_3,u_3,v_2,u_2)$ does not induce a $C_6$ in $G$, $uv_2 \in E$, and by the e.d.s.\ property, $uv'_2 \notin E$.  
Since $(u,v'_1,u'_2,v'_2,u'_3,v'_3)$ does not induce a $C_6$ in $G$, $uv'_3 \notin E$.
But then $(u,v_2,v_1,u_1,v'_1,u'_2,v'_2,u'_3,v'_3)$ (with midpoint $u$) induce an $S_{1,2,5}$ in $G$, which is a contradiction. 
Thus, $v_1$ and $v'_1$ do not have any common neighbor, and analogously, $v_1$ and $v'_3$, $v_3$ and $v'_1$, $v_3$ and $v'_3$ do not have any common neighbor.     

\medskip

Next assume that without loss of generality, $v_2u \in E$ and $v'_1u \in E$ with $u \notin V(P) \cup V(P')$. 
By the e.d.s.\ property, $uv'_2 \notin E$, and since $(u,v'_1,u'_2,v'_2,u'_3,v'_3)$ does not induce a $C_6$ in $G$, $uv'_3 \notin E$. 
But then $(v'_1,u'_1,u,v_2,u'_2,v'_2,u'_3,v'_3,u'_4)$ (with midpoint $v'_1$) induce an $S_{1,2,5}$ in $G$, which is a contradiction. 
Thus, $v_2$ and $v'_1$ do not have any common neighbor, and analogously, $v_2$ and $v'_3$, $v'_2$ and $v_1$, $v'_2$ and $v_3$ do not have any common neighbor.    

\medskip

Finally assume that without loss of generality, $u_2v \in E$ and $u'_2v \in E$ with $v \notin V(P) \cup V(P')$. 
Since $(u_2,v_2,v_1,u_1,v,u'_2,v'_2,u'_3,v'_3)$ (with midpoint $u_2$) does not induce an $S_{1,2,5}$ in $G$, $vu_1 \in E$ or $vu'_3 \in E$ but 
by the e.d.s.\ property, either $vu_1 \notin E$ or $vu'_4 \notin E$.

First assume that $vu_1 \in E$. Then by the e.d.s.\ property, $vu_4 \notin E$, $vu'_1 \notin E$ and $vu'_4 \notin E$. 
Since $(v,u_1,u_2,v_2,u'_2,v'_2,u'_3,v'_3,u'_4)$ (with midpoint $v$) does not induce an $S_{1,2,5}$ in $G$, 
$vu'_3 \in E$. 

Since $(v,u'_2,v'_1,u'_1,u'_3,v'_3,u'_4,u_2,v_2,u_3)$ (with midpoint $v$) does not induce an $S_{3,3,3}$ in $G$,
$vu_3 \in E$. But then $(v,u'_2,v'_1,u'_1,u'_3,v'_3,u'_4,u_3,v_3,u_4)$ (with midpoint $v$) induce an $S_{3,3,3}$ in $G$, which is a contradiction. 
Thus, $vu_1 \notin E$ and without loss of generality, $vu'_1 \notin E$.  

Next assume that $vu'_3 \in E$. Since $(v,u_2,v_1,u_1,u'_2,v'_1,u'_1,u'_3,v'_3,u'_4)$ (with midpoint $v$) does not induce an $S_{3,3,3}$ in $G$,
$vu'_4 \in E$, and by the e.d.s.\ property, $vu_4 \notin E$. 

Since $(u'_2,v'_2,v'_1,u'_1,v,u_2,v_2,u_3,v_3,u_4)$ (with midpoint $u'_2$) does not induce an $S_{1,2,5}$ in $G$, $vu_3 \in E$. 
But then $(v,u_2,v_1,u_1,u_3,v_3,u_4,u'_2,v'_1,u'_1)$ (with midpoint $v$) induce an $S_{3,3,3}$ in $G$, which is a contradiction.  
Thus, $u_2$ and $u'_2$ do not have any common neighbor, and analogously, $u_2$ and $u'_3$, $u_3$ and $u'_2$, $u_3$ and $u'_3$ do not have any common neighbor.     

\medskip

Thus, $dist_G(P,P') \ge 3$, and Lemma \ref{P7PP'nodist2} is shown.
\qed

\begin{lemma}\label{P7PP'distv2v'2=4} 
For $P_7$'s $P=(u_1,v_1,u_2,v_2,u_3,v_3,u_4)$, $P'=(u'_1,v'_1,u'_2,v'_2,u'_3,v'_3,u'_4)$ with midpoints $v_2,v'_2 \in D \cap Y$, $u_1,u'_1,u_4,u'_4 \in D \cap X$, $dist_G(v_2,v'_2)=4$.
\end{lemma}

\noindent
{\em Proof.}
Let $P=(u_1,v_1,u_2,v_2,u_3,v_3,u_4)$, $P'=(u'_1,v'_1,u'_2,v'_2,u'_3,v'_3,u'_4)$ induce $P_7$'s with midpoints $v_2,v'_2 \in D \cap Y$, 
$u_1,u'_1,u_4,u'_4 \in D \cap X$. By the e.d.s.\ property, $dist_G(v_2,v'_2) \ge 4$.

\medskip

Suppose to the contrary that $dist_G(v_2,v'_2) \neq 4$, say $dist_G(v_2,v'_2)=6$. Recall that by Lemma \ref{P7PP'nodist2}, $dist_G(P,P') \ge 3$.

\medskip

First assume that without loss of generality, $dist_G(u_2,u'_2)=4$, i.e., $(u_2,v,u,v',u'_2)$ induce a $P_5$ in $G$ with $v,u,v' \notin V(P) \cup V(P')$, 
and thus, $(v_2,u_2,v,u,v',u'_2,v'_2)$ induce a $P_7$ in $G$. 

If $u_1v \notin E$ then $(u_2,v_2,v_1,u_1,v,u,v',u'_2,v'_2)$ (with midpoint $u_2$) induce an $S_{1,2,5}$ in $G$, which is a contradiction.    
Thus, $u_1v \in E$. Moreover, if $u_3v \notin E$ then $(u_2,v_1,v_2,u_3,v,u,v',u'_2,v'_2)$ (with midpoint $u_2$) induce an $S_{1,2,5}$ in $G$, which is a contradiction. Thus, $u_3v \in E$. By the e.d.s.\ property, $vu_4 \notin E$. 

But then $(u_3,v_2,v_3,u_4,v,u,v',u'_2,v'_2)$ (with midpoint $u_3$) induce an $S_{1,2,5}$ in $G$, which is a contradiction.   
Thus, there is no $dist_G(u_2,u'_2)=4$, and analogously, there is no $dist_G(u_3,u'_3)=4$, $dist_G(u_2,u'_3)=4$, $dist_G(u_3,u'_2)=4$.

\medskip

Next assume that $(v_2,u,v,u',v',u'',v'_2)$ induce a $P_7$ in $G$ with $u,v,u',v',u'' \notin V(P) \cup V(P')$. 
If $uv_1 \notin E$ then $(v_2,u_3,u_2,v_1,u,v,u',v',u'')$ (with midpoint $v_2$) induce an $S_{1,2,5}$ in $G$, which is a contradiction.    
Thus, $uv_1 \in E$, and analogously, $uv_3 \in E$.  
But then $(u,v_2,v_1,u_1,v,u',v',u'',v'_2)$ (with midpoint $u$) induce an $S_{1,2,5}$ in $G$, which is a contradiction.

\medskip

Thus, $dist_G(v_2,v'_2)=4$, and Lemma \ref{P7PP'distv2v'2=4} is shown. 
\qed

\medskip

Recall that for at most two such $P_7$'s with midpoints in $D \cap Y$ and no $P_7$'s with midpoints in $D \cap X$, the e.d.s.\ problem is solvable in polynomial time. Now assume that there are three such $P_7$'s with midpoints in $D \cap Y$.

\begin{lemma}\label{P7sPP'P''atmostonedist3} 
For $P_7$'s $P=(u_1,v_1,u_2,v_2,u_3,v_3,u_4)$, $P'=(u'_1,v'_1,u'_2,v'_2,u'_3,v'_3,u'_4)$, $P''=(u''_1,v''_1,u''_2,v''_2,u''_3,v''_3,u''_4)$, with midpoints 
$v_2,v'_2,v''_2 \in D \cap Y$, $u_1,u'_1,u''_1,u_4,u'_4,u''_4 \in D \cap X$, 
there is at most one $dist_G(P,P')=3$. 
\end{lemma}

\noindent
{\em Proof.}
Let $P=(u_1,v_1,u_2,v_2,u_3,v_3,u_4)$, $P'=(u'_1,v'_1,u'_2,v'_2,u'_3,v'_3,u'_4)$, and $P''=(u''_1,v''_1,u''_2,v''_2,u''_3,v''_3,u''_4)$  induce $P_7$'s in $G$ with midpoints $v_2,v'_2,v''_2 \in D \cap Y$ and endpoints $u_1,u'_1,u''_1,u_4,u'_4,u''_4 \in D \cap X$. 
Recall that by Lemma \ref{P7PP'distv2v'2=4}, $dist_G(v_2,v'_2)=4$, $dist_G(v_2,v''_2)=4$, $dist_G(v'_2,v''_2)=4$, i.e., $3 \le dist_G(P,P') \le 4$, 
$3 \le dist_G(P,P'') \le 4$, $3 \le dist_G(P',P'') \le 4$. 

\medskip

Suppose to the contrary that without loss of generality, $dist_G(P,P')=3$ and $dist_G(P,P'')=3$. 
Since $dist_G(P,P')=3$, assume that $(v_2,u_2,v,u',v'_2)$ induce a $P_5$ in $G$. 
By the e.d.s.\ property, $vu_1 \notin E$ or $vu_4 \notin E$, say $vu_1 \notin E$. 

Since $(v'_2,u',v,u_2,u'_2,v'_1,u'_1,u'_3,v'_3,u'_4)$ (with midpoint $v'_2$) does not induce an $S_{3,3,3}$ in $G$, $u'v'_1 \in E$ or $u'v'_3 \in E$, 
say $u'v'_1 \in E$. 

Since $(u',v'_2,v'_1,u'_1,v,u_2,v_2,u_3,v_3)$ (with midpoint $u''$) does not induce an $S_{1,2,5}$ in $G$, 
$vu_3 \in E$.  
 
If $(v_2,u_2,v,u'',v''_2)$ induce a $P_5$ in $G$ then $(v,u_2,v_1,u_1,u',v'_1,u'_1,u'',v''_2,u''_3)$ (with midpoint $v$) induce an $S_{3,3,3}$ in $G$,  
which is a contradiction. Thus, for $u''v''_2 \in E$, $vu'' \notin E$, i.e., $(v_2,u_2,v',u'',v''_2)$ induce a $P_5$ in $G$ and $v'u' \notin E$ (else there is again an $S_{3,3,3}$ in $G$). 

Since $(v''_2,u''_2,v''_1,u''_1,u''_3,v''_3,u''_4,u'',v',u_2)$ (with midpoint $v''_2$) does not induce an $S_{3,3,3}$ in $G$, $u''v''_1 \in E$ or 
$u''v''_3 \in E$; without loss of generality, $u''v''_1 \in E$. 
 
But then $(u'',v''_2,v''_1,u''_1,v',u_2,v,u',v'_2)$ (with midpoint $u''$) induce an $S_{1,2,5}$ in $G$, which is a contradiction.
Thus, there is at most one $dist_G(P,P')=3$, i.e., 
$dist_G(P,P')=4$ and $dist_G(P,P'')=4$ or $dist_G(P,P')=4$ and $dist_G(P',P'')=4$ or $dist_G(P,P'')=4$ and $dist_G(P',P'')=4$.

\medskip

Then Lemma \ref{P7sPP'P''atmostonedist3} is shown.
\qed

\begin{lemma}\label{P7atmosttwoPP'} 
There are at most two such $P_7$'s with midpoints in $D \cap Y$. 
\end{lemma}

\noindent
{\em Proof.}
Let $P=(u_1,v_1,u_2,v_2,u_3,v_3,u_4)$, $P'=(u'_1,v'_1,u'_2,v'_2,u'_3,v'_3,u'_4)$ induce $P_7$'s in $G$ with midpoints 
$v_2,v'_2 \in D \cap Y$, $u_1,u'_1,u_4,u'_4 \in D \cap X$. Recall that $dist(v_2,v'_2)=4$. 

\medskip

Suppose to the contrary that there are three such $P_7$'s $P,P',P''$ with midpoints in $D \cap Y$, say $P''=(u''_1,v''_1,u''_2,v''_2,u''_3,v''_3,u''_4)$ with midpoint $v''_2 \in D \cap Y$, $u''_1,u''_4 \in D \cap X$. 

Recall that $dist(v_2,v'_2)=4$, $dist(v_2,v''_2)=4$, $dist(v'_2,v''_2)=4$ and by Lemma \ref{P7sPP'P''atmostonedist3}, there is at most one $dist_G(P',P'')=3$, 
say $dist_G(P,P')=4$ and $dist_G(P,P'')=4$.

\medskip

Assume that $(v_2,u,v,u',v'_2)$ induce a $P_5$ in $G$ with $u,v,u' \notin V(P) \cup V(P')$. 

If also $(v_2,u,v,u'',v''_2)$ induce a $P_5$ in $G$ then 
 $(v,u,v_2,u_2,u',v'_2,u'_2,u'',v''_2,u''_2)$ (with midpoint $v$) induce an $S_{3,3,3}$ in $G$, which is a contradiction.  
Thus, there is no such $P_5$ $(v_2,u,v,u'',v''_2)$ in $G$. 

Next assume that $(v_2,u,v',u'',v''_2)$ induce a $P_5$ in $G$. Then clearly $vu'' \notin E$. 
Since $(v''_2,u'',v',u,u''_2,v''_1,u''_1,u''_3,v''_3,u''_4)$ (with midpoint $v''_2$) does not induce an $S_{3,3,3}$ in $G$,
$u''v''_1 \in E$ or $u''v''_3 \in E$; without loss of generality, $u''v''_1 \in E$. 

But then $(u'',v''_2,v''_1,u''_1,v',u,v,u',v'_2)$ (with midpoint $u''$) induce an $S_{1,2,5}$ in $G$, which is a contradiction.
Thus, if $(v_2,u,v,u',v'_2)$ induce a $P_5$ in $G$ then $(v_2,u,v',u'',v''_2)$ does not induce a $P_5$ in $G$. 
But then there is again an $S_{1,2,5}$ in $G$, which is a contradiction. 

\medskip

Thus, there are at most two such $P_7$'s with midpoints in $D \cap Y$, and Lemma \ref{P7atmosttwoPP'} is shown.
\qed

\medskip

Then for at most two such $P_7$'s $P,P'$ with midpoints in $D \cap Y$, $P$ and $P'$ can be done in $G[N_0 \cup N_1]$. 
But then there are no $D$-vertices with midpoint of a $P_5$ in $G \setminus (N_0 \cup N_1)$, and by Section \ref{noDvertexmidpointP5}, the e.d.s.\ problem can be solved in polynomial time. Thus, Theorem \ref{EDS125S333frchordbippoltime} is shown. 

\medskip

\begin{footnotesize}

\end{footnotesize}

\end{document}